\documentclass[reprint,pre,aps,twocolumn,showpacs,longbibliography,floatfix,10pt,superscriptaddress]{revtex4-1}
\usepackage[dvips]{graphicx}
\usepackage{amsmath}
\usepackage{amssymb}
\usepackage[nice]{nicefrac}
\usepackage{color}
\usepackage{hyperref}
\usepackage{url}

\begin{document}

\title{Formal Fluctuation-Response Relations for Non-Stationary Systems: The Dynamic Conjugate Variable}

\author{Igor M. Sokolov}
\email{Corresponding author:\\ igor.sokolov@physik.hu-berlin.de}
 \affiliation{Institut f{\"u}r Physik, Humboldt-Universit{\"a}t zu Berlin, Newtonstra{\ss}e 15, D-12489 Berlin, Germany}

\begin{abstract}
Fluctuation-response relations (FRRs) connect the linear response of a system to an external perturbation with properties of spontaneous fluctuations in the unperturbed system. 
We provide a simple derivation of FRRs for non-stationary dynamics following an almost standard way based on introducing a variable conjugated to perturbation (a dynamic conjugate). The derivation 
relies solely on the Markovianity of the underlying dynamics, with the only additional assumption that the transition PDF of the process is differentiable with respect to the strength of a constant,
time-independent perturbation. 
The structure of this conjugate variable is however unusual: it is a two-time one, and reduces to a usual single-time conjugate in stationary situations. 
We show how several known results follow from this approach. 
\end{abstract}

\maketitle

\section{Introduction.}
Fluctuation-response relations (FRRs), or fluctuation-dissipation relations, or theorems (FDTs) connect the (linear) response of a system to an external perturbation with properties 
of spontaneous fluctuations in the unperturbed system. We will continuously use the first naming, since in many cases the FRRs are applied to situations outside physics, where the mere
notion of dissipation is not defined, like in \cite{Sato,Yan,Droste,Lindner}. The FRRs comprise different types of relations: The initial discussion of  Kubo \cite{Kubo} 
distinguished between the FRRs of the first and of the second kind. Ref.~\cite{Villamaina} counts three kinds of FRRs. The FRRs, as we consider them here, have to be distinguished from the 
Kubo-like formulae for kinetic coefficients, giving a rougher kind of description, and from fluctuation theorems (see e.g. \cite{Sevick} for a review) giving a finer one.  A comprehensive historical 
review can be found in \cite{Darrigol}, 
and a relatively exhaustive discussion of FRRs in a classical setting is given in Ref.~\cite{Vulpiani}. Some newer developments are mentioned in \cite{Ewa} and in references therein. Many non-equilibrium 
applications are discussed in \cite{Baldovin,Ewa_new}. Since the literure on the topic is immence, we don't aim on covering all types of relations, and all different approaches to
the problem, but mention several additional important papers, \cite{Sarracino,Seifert_2}, and especially \cite{Baiesi}, containing specific discussion of physical systems far from equilibrium.

The history of FRRs started from investigation of thermodynamic systems close to equilibrium \cite{Darrigol}, where 
the FRRs (of the first kind) are typically applied for calculation of the linear response functions or susceptibilities, the time integrals thereof, leading to a possibility to express 
responses via equilibrium properties of the system.
Out of equilibrium, the standard application of FRR  needs for a knowledge of a sufficient Markovian embedding for the system's dynamics, see e.g. \cite{Goerlich}. This motivates a different, ``off-label'' use of FRRs: 
Under particular conditions, the fulfillment or non-fulfillment of FRR allows to answer the Onsager and Machlup's question \cite{OMa}: ``How do you know you have taken enough variables (enough
measurements on the system) for it to be Markoffian?'' This problem is discussed e.g. in \cite{Villamaina,Willareth} and \cite{Engbring} for linear and non-linear cases, respectively.  
We note that entering the non-linear response regime does not considerably improve the sensitivity of the approach, and may on the contrary deteriorate it~\cite{Podhaisky}. 
This is one of the reasons why we stick to a linear response regime here. In view of this application, in the present discussion we only consider the classical setting. 
This allows us not to discuss peculiarities of multitime measurements which may be necessary to experimentally assess the validity of such relations. Thus, we assume that the measurement instrument 
provides a ``continuous record,'' as mentioned by Onsager and Machlup \cite{OMa}. For the sake of concreteness we consider systems with continuous state variables and continuous dynamics. 

Another reason to focus on the linear response is its universality: Knowing the response to one specific shape of the perturbation, typically a $\delta$-pulse, or 
a $\theta$-shaped perturbation switched on or off, one knows the response to any other perturbation, due to the superposition property. In the nonlinear case, 
one typically restricts either the kind of the response (say, linear-quadratic, or linear-cubic), or the type of signal (say, a periodic one), as it is often done in nonlinear optics. 
These nonlinear cases stay outside of our discussion here.

Let us proceed by introducing some notions and  terminology which will be continuously used in what follows.

Ref.~\cite{Caprini} distinguishes between the two classes of approaches to FRRs (for stationary cases, including the equilibrium ones): ``Class A requires the knowledge of the stationary distribution, while class B only requires
the knowledge of the microscopic dynamical rules.'' A typical discussion in the class A approach starts from the definition of a variable conjugated to perturbation: 
This variable describes relative change in the stationary distribution when the perturbation is applied.
In the class B approaches a conjugate is obtained without invoking the stationary distribution. For important examples of the class B approach one can refer to \cite{Seifert}, which also stresses 
the existence of different kinds of conjugates to the same forcing, all corresponding to the same behavior of relevant means.
We will call FRRs obtained via approaches akin to class A ones \textit{formal FRRs}: They represent universal relations connecting the response with probability distributions, but
for practical applications and for physical interpretation of results these distributions have still to be obtained theoretically from micro- or mesoscopic rules, or measured in an experiment. 
The only exclusion are the cases close to equilibrium, where the properties of the stationary (equilibrium) distribution follow immediately from the canonical distribution function, of from thermodynamic theory
of fluctuations, see Sec. \ref{sec:Bilinear}.

In the present work we consider formal FRRs in general, not necessarily stationary situations. In some sense, our generalization follows the idea of approaches based on transition rates mentioned in \cite{Baldovin}, but the realization of this idea is different, following the class A, and not the class B route.
As a reference, the corresponding class A approach for the stationary case is discussed in Sec. \ref{sec:Stand} in exactly the same notation that is 
used in the following discussion of non-stationary cases. The new twist in our following discussion, Sec.~\ref{sec:Main}, is that the same steps are applied not to the stationary single-time probability density function (PDF), 
but to time-dependent \textit{transition} PDFs. 
Our generalization is based on a minimal  number of assumptions, and represents a prototypical form of several modified FRRs discussed recently, thus providing a unified
mathematical basis of such discussions. In the stationary case for a system showing mixing dynamics our formal FRR turns out to be a customary FRR of the first kind.  

Our discussion is illustrated by several examples.  These examples are, on purpose, chosen to be simple, and to correspond to well-understood physical situations. 
The examples illustrate how the approach works, and allow to put questions giving the direction of further inversigations. 
The non-trivial physical situations, also in cases going outside of our setting here, will be considered elsewhere.

\section{Preliminaries}

\subsection{Problem setting}

We consider linear response of a physical (biological, socio-economical) system S to an external perturbation, and address the 
following experimental situation: A system S, prepared at time $t_0$ by some well-prescribed procedure (defining the ensemble $\{ \mathrm{S} \}$ of systems at preparation), 
is considered as a black box with several tuning knobs defining the vector $\vec{F}$  of external parameters with components $F_\alpha$. The system is subjected to the changes of these parameters 
following some protocol $\vec{F}(t)$. These changes, which in what follows will be called ``forcing'' or ``force,'' independent on their physical nature, influence the internal dynamics of the system. 
The dynamics of the state variables $\mathbf{x}(t)= \{x_i(t)\}$ describing the evolution of the system is assumed Markovian, so that the knowledge of initial conditions for the internal variables 
at some time $t'$ allows for prediction of the probability distribution of 
these variables at any later instant of time $t>t'$ provided the protocol $\{\vec{F}(t)\}$ of $\vec{F}$ is known. Measured is some specific observable, the \textit{variable of interest} (VoI) 
$\mathcal{O}$. The value of $\mathcal{O}(t) \equiv \mathcal{O}(\mathbf{x}(t),t)$ only depends on the values of $\mathbf{x}(t)$ at the instant of measurement (``single-time VoI''). 
In our main discussion we will keep the scalar notation, assuming $F$ and $\mathcal{O}$
to have only one component each. We will concentrate on the observables $\mathcal{O}(\mathbf{x}(t))$ which are not explicitly time-dependent. The generalization to 
explicitly time-dependent ones is simple.
The formulas for multicomponent variables (useful e.g. for obtaining Curie principles) also follow by a trivial generalization.

We note that in the equilibrium or stationary cases different types of perturbations can be considered leading to 
essentially the same class of FRRs. Thus, instead of perturbing external parameters $\vec{F}$ (different from the state variables) one might perturb the value of 
$\mathbf{x}$ at some instant of time, and then consider how the system returns to equilibrium or stationary state (see e.g. the discussion in \cite{Sarracino}). 
One can call the corresponding relations for perturbed $\mathbf{x}$ the regression relations (with the Onsager's 
regression hypothesis being one of such), and to reserve the name ``fluctuation-response relations'' for perturbing $\vec{F}$.  
In classical systems close to equilibrium the two situations are closely related: the thermodynamic coordinate,
and the conjugate thermodynamic force come in pairs, and in involution: they are mutually conjugate. This property may be lost in nonstationary systems. 
Beyond stationarity, there is a reason to distinguish between the two situations, and in what follows we concentrate on the first one, assuming that some parameters of the system different
from components of the state vector are perturbed. 

A Markov process is defined as a ``memoryless'' process for which the conditional probability density $p(\mathbf{x}_N,t_N|\mathbf{x}_{N-1},t_{N-1}; \mathbf{x}_{N-2},t_{N-2}; ... ; \mathbf{x}_1,t_1)$ with
$t_N \geq t_{N-1} \geq t_{N-2} \geq ... \geq t_1$ is equal to $p(\mathbf{x}_N,t_N|\mathbf{x}_{N-1},t_{N-1})$, i.e. is fully determined by the values of $\mathbf{x}$ at the previous instant of time, but not at preceding ones.
The two relations following from this assumption,
\begin{equation}
p(\mathbf{x}_2,t_2) = \int p(\mathbf{x}_2,t_2|\mathbf{x}_1,t_1)p(\mathbf{x}_1,t_1) d \mathbf{x}_1 
 \label{eq:CInt}
\end{equation}
and the Chapman-Kolmogorov equation
\begin{equation}
p(\mathbf{x}_2,t_2|\mathbf{x}_1,t_1) = \int p(\mathbf{x}_3,t_3|\mathbf{x}_2,t_2) p(\mathbf{x}_2,t_2|\mathbf{x}_1,t_1) d \mathbf{x}_2
\label{eq:CK0}
\end{equation}
(\cite{van_Kampen}, Chapter IV, see also Refs.~\cite{Feller_ex,Canturk} for the discussion of some mathematical aspects) build the basis of our discussion. 

Typically, the transition PDF is considered to correspond to a proper conditional probability distribution. If the system possesses absorbing states, and these are included 
into the state space of the system, our description applies to the systems with absorbing states as well. These, however, do not have to be included into the state space, and all
our probabilities / densities may be defined only for transient states, and be non-proper ones, provided Eqs.~(\ref{eq:CInt}) and (\ref{eq:CK0}) hold for these non-proper probabilities / densities
for all times, see Sec.~(\ref{sec:Nonprop}).

\subsection{Linear response} 

The response of a system close to equilibrium or other \textit{stable} stationary state to the external forcing describes the deviation of the ensemble mean value $\langle \mathcal{O}(t) | \{F\} \rangle$
of the VoI under forcing from the value $\langle \mathcal{O} |0 \rangle$ corresponding to the reference state in the absence thereof. The both mean values are assumed to exist, which defines the 
domain of applicability of the corresponding relations (i.e. the corresponding class of VoIs). 
The condition in the first mean corresponds to the whole protocol of $F(t')$ for $t_0 < t' \leq t$, with $t_0$ being the preparation time. 
The dynamics of the system in the absence of the external forcing or under the action of the constant force is assumed time-invariant, and the stable stationary state 
(if it exists) is reached spontaneously at longer times. Since at $t_0$ the system is in a stationary state, the second mean, corresponding to the reference state,
is time-independent. The preparation procedure which corresponds to this situation is to create the system at an earlier time, let it reach the equilibrium or the stable 
stationary state, and choose $t_0$ such, that at $t_0$ the equilibrium or stationarity is reached.
Then the linear response assumes that 
\[
 \langle \mathcal{O}(t) | \{F\} \rangle - \langle \mathcal{O} |0 \rangle = \int_{t_0}^t \chi(t,t') F(t') dt'.
\]
Since at $t_0$ the system is already in the stationary state, and the forcing vanishes for $t' < t_0$, one can take $t_0 \to - \infty$. 
Here, $\chi(t,t')$ is the response function of the system, which does not depend on the forcing (otherwise the response will be nonlinear).
Therefore, the properties of $\chi(t,t')$ are governed by the system's behavior in the stationary state in the absence of the forcing only. 
The time-homogeneity of the stationary state then implies that $\chi(t,t')$ depends only on the difference of its arguments, so that we may write
\[
 \langle \mathcal{O}(t) | \{F\} \rangle - \langle \mathcal{O} |0 \rangle = \int_{-\infty}^t \chi(t - t') F(t') dt',
\]
a convolution form, well known e.g. in the electrodynamics of media. Note that the same discussion applies to the behavior of a system prepared
under the action of a constant force $F$, which may then change in an experiment starting from the preparation time. We note again that the stability of the stationary
state, now in the presence of the force $F$, is assumed.

Now let us assume that the stable stationary state is reached both in the absence and in the presence of a constant force and consider the following situation: 
The system is prepared in the absence of the force, and the constant force $F$ is switched on at time $t'$. 
In this case, at long times $t \to \infty$ we have
\begin{equation}
 \langle \mathcal{O}(t \to \infty) | F \rangle - \langle \mathcal{O} |0 \rangle = F \int_{t'}^\infty \chi(t - t') dt'.
 \label{eq:StaLR}
\end{equation}
The time integral on the right hand side then represents the static susceptibility of the system: $\mathcal{X}_{\mathcal{O}F} = \int_0^\infty \chi(\Delta t) d \Delta t$
which is finite provided the stationary states both in the presence and in the absence of the force exist and are stable. 
In this case we have: $\langle \mathcal{O}| F \rangle - \langle \mathcal{O} |0 \rangle = \mathcal{X}_{\mathcal{O}F} F$.

Passing to nonstationary situations introduces the following changes into the linear response description.
First, the reference state is no more a stationary one, so that the perturbation induces the deviation from an explicitly time-dependent 
behavior of $\mathcal{O}(t)$ in the absence of the force. Second, for a system prepared in a nonstationary state at time $t_0$, the response function
explicitly depends on three times, $t$, $t'$ and $t_0$.

\color{black}

Thus, the linear response of the system to the force leads to deviation of the observed dynamics $\langle \mathcal{O}(t) |t_0, \{F\} \rangle$ under forcing from the time-dependent
behavior of $\langle \mathcal{O}(t) | t_0,  0 \rangle $ in an unperturbed system, 
constituting a reference state: 
\begin{equation}
\langle \mathcal{O}(t) |t_0, \{F\} \rangle - \langle \mathcal{O}(t) | t_0,  0 \rangle = \int_{t_0}^t \chi(t,t',t_0) F(t') dt',
\label{eq:LR2}
\end{equation}
where the condition in the first mean again corresponds to the whole protocol of $F(t')$ for $t_0 < t' \leq t$. The function $ \chi(t,t',t_0)$ is the linear response function of our system.  
We will assume that the linear response regime does exist, i.e., that Eq.~(\ref{eq:LR2}) is valid with sufficient accuracy provided the forcing $F(t)$ is weak enough at all times. 
Note, however, that for some specific VoIs the function $ \chi(t,t',t_0)$ might vanish identically, while for other ones it is non-zero. This is a situation corresponding to 
a Curie principle, and will be discussed in one of our examples.

We note that the ensemble $\{\mathrm{S}\}$ of systems at preparation, i.e., the initial condition
giving the PDF $p(\mathbf{x}_0|t_0)$, is assumed to be known, and is the same, whenever the system is prepared, i.e. whatever $t_0$ is chosen to start the experiment. 
For a given ensemble at preparation, and since the time by itself is homogeneous, $\chi(t,t',t_0)$ is a function of only two independent arguments (time lags, or ages). 
Typically, one chooses the age of the system at $t'$, $t_a = t'-t_0$ and the time lag $\Delta t = t-t'$ as arguments. Depending on our aims, we will use
both notations (the two- and the three-time one), depending on which one simplifies the derivations and discussions. 

The limits of integration in Eq.~(\ref{eq:LR2}) correspond to
\textit{two} causality conditions. The first is a usual one: There is no response to the forcing values after the measurement
time $t$. The second one states that since $\{ \mathrm{S} \}$ was prepared at $t_0$ in a given state,
its further evolution does not depend on what happened before preparation: this condition is an immediate consequence of the Markovian property. 

To find the response function, it is enough to consider some specific protocol of perturbation. Considering a perturbation as a short pulse
allows to calculate the response to any protocol by considering the integral in Eq.~(\ref{eq:LR2}) as a Riemann one. Taking a limit of a $\delta$-pulse leads to the most common definition of 
$\chi$ via the functional derivative w.r.t. perturbation, $ \chi(t,t',t_0) = \frac{\delta \langle \mathcal{O}(t) | t_0, F(t') \rangle}{\delta F(t')}$.
On the other hand, one can consider the response to switching on and off the constant forcing,
and use the Lebesgue definition of the integral. In particular, we will consider two protocols: Switching a weak force on at time $t'$ ($t_0 < t' \leq t$), $F_+(t) = F \Theta(t-t')$, and switching it off: $F_-(t) = F \Theta(t'-t)$. 
The corresponding cases will be called the $F_+$ and $F_-$ setups in what follows. In equilibrium, the discussion of the $F_-$ setup belongs to scientific 
folklore, and can be even found in the Wikipedia \cite{Wiki}. The proof for the classical case there goes back to Lenk, \cite{Lenk}. Ref.~\cite{Vulpiani} attributes the $F_-$ setup to Onsager,
but it is hard to point out, where exactly the setup is discussed. The earliest explicit mentioning of the $F_-$ setup is probably by Greene and Callen, \cite{GreeneCallen1}. 
The $F_+$ approach is older, and was first applied by Takahasi \cite{Takahasi} for after-effect functions. 

Within the $F_+$ and $F_-$ setups we thus have:
\begin{eqnarray}
&& \! \!\! \!  \langle \mathcal{O}(t) |t_0, F_+ \rangle - \langle \mathcal{O}(t) | t_0,  0 \rangle = F \int_{t'}^t \chi(t,t'',t_0) dt'',
\label{eq:LRFplus} \\
&& \! \!\! \!  \langle \mathcal{O}(t) |t_0, F_- \rangle - \langle \mathcal{O}(t) | t_0,  0 \rangle = F \int_{t_0}^{t'} \chi(t,t'',t_0) dt''.
\label{eq:LRFminus}
\end{eqnarray}
This gives us the response functions
\begin{equation}
\chi_\pm(t,t',t_0) = \mp \frac{1}{F} \frac{d}{dt'} [\langle \mathcal{O}(t) | t_0, F_\pm \rangle- \langle \mathcal{O}(t) | t_0, 0 \rangle]  .
\label{eq:2t}
\end{equation}
Later we will show that $\chi_{\pm}(t,t',t_0)$ coincide, $\chi_{\pm}(t,t',t_0) = \chi(t,t',t_0)$, as they should. Knowing $\chi(t,t',t_0)$ 
one can calculate the linear responce to arbitrary forcing using Eq.~(\ref{eq:LR2}).

For a system tending to a stationary state at longer times, $\chi(t,t',t_0)$  tends to the time-homogeneous one when $t_0$ tends to $-\infty$ but the difference $\Delta t = t-t'$ is kept constant,
$\lim_{t_0 \to -\infty} \chi(t,t',t_0) = \chi(t-t')$:
The time-dependence of $\chi$ reduces to the one on the time lag $t-t'$. Then the standard linear response relation, Eq.~(\ref{eq:StaLR}) applies.

\subsection{An additional note}

Before going further with our discussion, we make the following important note. 
The approach we outline in what follows is discussed for single-time VoIs only. Let us assume that we know, how many variables do we have to consider to have
a Markovian description of the dynamics of our system. We will say that these variables constitute a \textit{minimal list} of relevant variables. The description via the 
minimal list (i.e. the Markovian embedding) of our system allows obtaining FRR and response functions for all VoIs which are functions of the variables of a minimal list.
We may however be interested in a behavior of an observable which is not a function of these variables. For example, the velocity $\mathbf{v}(t)$ of a massive colloidal particle 
in a fluid medium (see Sec. \ref{Sec:VelOU}) is described by the Markovian Ornstein-Uhlenbeck process. The observable we are interested in may however be the particle's position $\mathbf{x}(t)$, 
(essentially, the displacement $\Delta \mathbf{x}(t) = \mathbf{x}(t) - \mathbf{x}(t_0)$) which is a linear functional (time integral)
of this velocity. This observable is not a single-time VoI with respect to the variables of the minimal list. Considering the augmented state space of the system (which in our example is the 
phase space of vectors containing velocities and coordinates, see the example in Sec. \ref{Sec:AugOU}), we get a Markovian description of the evolution of these state vectors, constituting the 
\textit{augmented list} of 
Markovian variables in which the VoI is a single-time one, so that all our following discussion applies to this case as well. 
Therefore, the kind of Markovian embedding necessary to obtain the response of a given VoI to a given perturbation is not only the property of the system, but a property of the experimental
situation, and its specification should include the discussion of both the system and the VoI. In the present work, we will restrict this discussion to one simple example only,
which shows that passing to an augmented state space may be of advantage even if discussing VoIs not involving these additional variables.
A thorough investigation of this situation is left for forthcoming work.

\section{A reference (class A) approach for stationary states\label{sec:Stand}}

Before turning to non-stationary cases, and to set the stage, we reproduce the standard class A derivation, where we closely follow Ref.~(\cite{Goerlich}). This is essentially the
same derivation as is discussed in the Supplemental Material to Ref.~\cite{Sokolov}, but now we use the notation which is consistent with the rest of the work, and allows generalization 
to non-stationary cases. The FRR which we derive in this way is related to the FDT of Prost et al \cite{Prost} in NESS, and is a standard FDT 
of the first kind in the equilibrium situation. The corresponding derivation uses the $F_-$ setup. 

We  assume that under the action of any admissible $F$ the system at $t'$, before the force has been switched off, has achieved a stationary state characterized by the 
PDF $p(\mathbf{x}'|F)$. The time $t'$ may be set to zero, which is done in what follows. We now have
\[
\langle \mathcal{O}(t)| F_- \rangle = \int \! \!\! \! \int d\mathbf{x} d\mathbf{x}' \mathcal{O}(\mathbf{x}) p(\mathbf{x},t|\mathbf{x}',0) p(\mathbf{x}'|F).
\]
Therefore,
\begin{eqnarray*}
&& \langle \mathcal{O}(t)| F_- \rangle - \langle \mathcal{O}(t)| 0 \rangle = \int  \! \!\! \!  \int d\mathbf{x} d\mathbf{x}' \mathcal{O}(\mathbf{x}) p(\mathbf{x},t|\mathbf{x}',0)\\
&& \qquad \times  [p(\mathbf{x}'|F) - p(\mathbf{x}'|0)].
\end{eqnarray*}
Introducing 
\[
 \widetilde{Y}(\mathbf{x}, F) = \frac{p(\mathbf{x}|F) - p(\mathbf{x}|0)}{p(\mathbf{x}|0)}
\]
we thus may write 
\begin{eqnarray*}
 && \langle \mathcal{O}(t)| F_- \rangle - \langle \mathcal{O}(t)| 0 \rangle =  \\
 && \qquad \int  \! \!\! \!  \int d\mathbf{x} d\mathbf{x}'  \mathcal{O}(\mathbf{x}) \widetilde{Y}(\mathbf{x}',F) p(\mathbf{x},t|\mathbf{x}',0) p(\mathbf{x}'|0) = \\
 && \qquad \int  \! \!\! \!  \int d\mathbf{x} d\mathbf{x}'  \mathcal{O}(\mathbf{x}) \widetilde{Y}(\mathbf{x}',F) p(\mathbf{x},t;\mathbf{x}',0) \\
\end{eqnarray*}
where $p(\mathbf{x},t;\mathbf{x}',0)$ is the joint PDF of internal variables at times $t$ and $0$ for $F=0$. Thus,
\begin{equation}
\langle \mathcal{O}(t)| F_- \rangle - \langle \mathcal{O}(t)| 0 \rangle =  \left\langle  \mathcal{O}(t) \widetilde{Y}(F,0) \right\rangle_0 .
 \label{eq:FDTL}
\end{equation}
By writing $ \widetilde{Y}(F, 0)$ we stress that the variable  $\widetilde{Y}$ (the nonlinear conjugate to the force) is calculated or measured at $t=0$. The subscript 0 in the last 
mean indicates that this one is calculated using the joint PDF in the absence of the force. 
Eq.~(\ref{eq:FDTL}) is the nonlinear fluctuation-dissipation theorem of Ref.~\cite{Engbring}.
Now, let us assume that $p(\mathbf{x}|F)$ is differentiable w.r.t. $F$, and approximate $ \widetilde{Y}(\mathbf{x},F)$ in the linear order:
\begin{equation}
  \widetilde{Y}(\mathbf{x}, F) \simeq F \left. \nabla_F \ln p(\mathbf{x},F) \right|_{F=0} \equiv F Y(\mathbf{x})
  \label{eq:LinAp}
\end{equation}
with $Y(\mathbf{x}) = \left. \nabla_F \ln p(\mathbf{x},F) \right|_{F=0}$ being the linear \textit{static} conjugate variable. 
The dimension of the conjugate variable is an inverse of the dimension of the force $F$.
We note that since $\widetilde{Y}(\mathbf{x}, F)$ and therefore $Y(\mathbf{x})$ follow from the stationary $p(\mathbf{x}|F)$ they are 
functions of state variables only, and do not explicitly depend on time.
From Eqs.~(\ref{eq:FDTL}) and (\ref{eq:LinAp}) we get:
\begin{equation}
\langle \mathcal{O}(t)| F_- \rangle - \langle \mathcal{O}(t)| 0 \rangle =  F \left\langle  \mathcal{O}(t) Y(0) \right\rangle_0 .
 \label{eq:FDTStan}
\end{equation}
Eq.~(\ref{eq:FDTStan}) is the customary FRR in a stationary state (equilibrium or not).

We stress that the only assumption except for Markovianity which is necessary for the derivation above is the differentiability of $p(\mathbf{x},F)$ w.r.t. to the forcing, Eq.~(\ref{eq:LinAp}).
We will say that the linear response domain does exist, whenever Eq.~(\ref{eq:LinAp}) holds. The linear response \textit{of the system} to the forcing does not exist only if the corresponding 
derivative diverges or vanishes, so that the dependence of $\widetilde{Y}(\mathbf{x}, F) $ on $F$ is either singular or ``flatter'' than linear. Such cases (encountered e.g. at points of 
phase transitions) correspond to genuine absence of the linear response regime. If the linear response of the system to the forcing exists, Eq.~(\ref{eq:FDTStan}) should hold at least for 
$\mathcal{O}(\mathbf{x}) = Y(\mathbf{x})$. 

For $\mathcal{O}(\mathbf{x}) = Y(\mathbf{x})$, Eq.~(\ref{eq:FDTStan}) represents the generalized FDT of Prost et al \cite{Prost}
(up to a different sign convention). Thus, this FRR has a purely probabilistic nature, and does not have to be derived from a finer fluctuation relation of Ref. \cite{Sasa}. 

On the other hand, even if $Y(\mathbf{x})$ is well-defined, the integral on the r.h.s. of Eq.~(\ref{eq:FDTStan})
might vanish, say on symmetry reasons. In this case we will say, that the linear response \textit{of a particular observable} to the forcing does exist but the corresponding 
response function vanishes. Examples for such
situations are numerous. One of them will be considered in Sec.~\ref{sec:Trap}. We note that the same discussion applies also to our generalization of the FRR to non-stationary situations given in 
Sec.~\ref{sec:Main}.

Now we consider three examples of applying the relation Eq.~(\ref{eq:FDTStan}).

\subsection{Example 1: An equilibrium case with bilinear coupling \label{sec:Bilinear}}

In equilibrium, the relation 
\[
 \langle Y(t)| F_- \rangle - \langle Y(t)| 0 \rangle =  F \left\langle Y(t) Y(0) \right\rangle_0
\]
it is the standard FDT of the first kind. In the canonical case with bilinear coupling of a force to a single coordinate $x$, we have $p(x|F) = Z^{-1} \exp\{-\frac{1}{k_B T} [U(x) - Fx] \}$
with $U(x)$ being the potential energy in the unperturbed Hamiltonian, and get
\begin{equation}
 Y = \frac{x}{k_B T}
 \label{eq:Ystand}
\end{equation}
provided the dependence of the partition function $Z$ on $F$ appears only in a higher order in $F$, which is the case for any non-singular $U(x)$, for example, for a quadratic potential. 
The same is the case if we use the standard Onsager-like 
approach close to thermodynamic equilibrium and write $p(\Delta X) \propto \exp[\Delta S(\Delta X)/k_B]$ with $X$ being an extensive thermodynamic variable, $\Delta X = X - X_{\mathrm{eq}}$, and $S$ is
the entropy of the isolated composite system (comprising the system of interest and the bath), having a simple quadratic maximum at the equilibrium value of $X_{\mathrm{eq}}$. In this case 
$Y(\Delta X) = \frac{1}{k_B} \left. \frac{\partial \Delta S}{\partial X} \right|_{X=X_{\mathrm{eq}}+\Delta X}$, and is proportional to the change $\Delta x$ in the intensive thermodynamic variable $x$ conjugated to $X$.  
In both cases, the perturbation ($F$ or $\Delta X$), and the conjugate, being the derivative of a scalar function with respect to the former, come in involution, and are 
connected via the Legendre transformation. This also explains why the approaches based on the forces and on considering the after-effects of abrupt changes in coordinates (i.e., the regression 
relations) are essentially equivalent. A deeper reason for such an equivalence of regression and fluctuation-response relations is that both kinds of perturbations essentially 
manipulate the initial condition for the further free dynamics, and not the dynamics itself.
This structure of mutually conjugated pairs of variables may be partly conserved in nonequilibrium generalizations of FDTs to NESS, see the discussion in 
\cite{Altaner},
but is typically lost in our non-stationary generalization, since the coordinates in the transition probability densities come twice and at different times, \textit{vide infra}. 
In this case our generalized response relation, and regression relations lose the immediate connection to each other, and we only concentrate on the former.  

\subsection{Example 2. A particle in an an optical trap. \label{sec:Trap}}

As a second example we consider the situation discussed in \cite{Goerlich}.
Namely, we consider an overdamped harmonic oscillator (Ornstein-Uhlenbeck process) in equilibrium, but the perturbation applied is not
an external force but a change in the corresponding spring constant. The system is prepared with the spring constant $k=k_0 + \Delta k$ and let equilibrate. The perturbation $\Delta k$ is then switched off 
at $t=0$. The experimental realization of this system is considered in \cite{Goerlich,Goerlich2} and corresponds to a Brownian particle in an an optical trap under the 
change of laser intensity. In this case we have
\[
 p(x|k) = \sqrt{\frac{k}{2 \pi k_B T}} \exp \left(- \frac{k x^2}{2 k_B T} \right), 
\]
and 
\[
Y = \frac{1}{2 k_0} - \frac{x^2}{2 k_B T}.
\]
Note that the result does not change in the isotropic situation if $x$ were a vector. 
We thus have 
\[
 \langle \mathcal{O}(t)| \Delta k \rangle - \langle \mathcal{O}(t)| 0 \rangle =  \Delta k \left\langle  \mathcal{O}(t) Y(0) \right\rangle_0 .
\]
The reason to discuss this example is as follows. The conjugate variable $Y$ defines an immediate response of the system to a perturbation of the spring's stiffness.
The standard FDT which ensues by taking $\mathcal{O}(x) = Y(x)$ will give us 
\begin{eqnarray*}
&& \langle Y(t)| \Delta k \rangle - \langle Y(t)| 0 \rangle =  \Delta k \left\langle  Y(t) Y(0) \right\rangle_0 \\
&&= \Delta k \left(\frac{1}{4 k_0^2} - \frac{1}{2k_0} \frac{\langle x^2(t) \rangle}{2 k_B T} - \frac{1}{2k_0} \frac{\langle x^2(0) \rangle}{2 k_B T} + \frac{\langle x^2(t) x^2(0) \rangle}{(2 k_B T)^2}\right).
\end{eqnarray*}
Using the equipartition theorem, $\frac{k_0 \langle x^2 \rangle}{2} = \frac{k_b T}{2}$, the result can be rewritten as
\[
  \langle Y(t)| \Delta k \rangle - \langle Y(t)| 0 \rangle = \Delta k \left( \frac{\langle x^2(t) x^2(0) \rangle}{(2 k_B T)^2} - \frac{1}{4 k_0^2} \right),
\]
where by applying the Wick's theorem one can express the numerator in the first term via $C_{xx}^2$ \cite{Goerlich}. The static conjugate $Y$ shows the non-vanishing linear response
to changes in $k$. On the other hand, assuming $\mathcal{O}(x) = x$, we see that 
the response function vanishes identically, since all odd moments of a symmetric Gaussian are zero: 
\[
 \left\langle  x(t) Y(0) \right\rangle_0 = \frac{\langle x \rangle}{2 k_0} - \frac{\langle x(t) x^2(0)\rangle}{2 k_B T} = 0. 
\]
This is a simplest case of a Curie principle, since $x$ behaves as a vector, and 
$k$ is a scalar. The example stresses that the variable conjugated to a weak forcing does not have to be linear
in the internal variables of the system, even if the system is a linear one. It also shows that the knowledge of the conjugate variable is of independent value since it shows what 
variables have a nontrivial response function to a given perturbation, and what ones have the vanishing one. We also see how the symmetry considerations enter the game. 
Note that Ref.~\cite{Goerlich2} also considers nonlinear response (of $x^2$) to exactly the same perturbation.

After discussing these essentially equilibrium examples (a textbook and a newer one) we turn to a NESS case, and consider the response of a Brownian motion under Poissonian resetting \cite{EvaMaj}.
The response of the process to the external force $F$ is considered in detail in Ref.~\cite{Sokolov}, and here we only concentrate on the response to changing the resetting rate $\lambda$. 

\subsection{Example 3. Brownian motion under Poissonian resetting.}

The PDF of a reset Brownian motion under Poissonian resetting with rate $\lambda$ is given by \cite{EvaMaj}
\begin{equation}
 p(x; \lambda) = \frac{1}{2} \sqrt{\frac{\lambda}{D}} \exp \left(- \sqrt{\frac{\lambda}{D}} |x| \right),
 \label{eq:EM}
\end{equation}
a Laplace distribution. The variance of this distribution is 
\[
 \langle x^2 \rangle = 2 \frac{D}{\lambda},
\]
and the absolute first moment reads:
\[
 \langle |x| \rangle = \sqrt{\frac{D}{\lambda}}.
\]

Let us consider a small change of the resetting rate: $\lambda \to \lambda + \Delta \lambda$.
The static conjugate is 
\begin{eqnarray*}
 Y &=& \lim_{\Delta \lambda \to 0} \frac{1}{p(x;\lambda)} \frac{\partial}{\partial \Delta \lambda} p(x;\lambda + \Delta \lambda) \\
 &=& \frac{\partial}{\partial \lambda} \ln p(x;\lambda) = \frac{1}{2 \lambda}-\frac{|x|}{2 \sqrt{\lambda D}} .
\end{eqnarray*}
The conjugate has a dimension of the time and is spatially symmetric, and thus, like in our previous case, the response of all
odd spatial VoIs to our perturbation vanishes. Taking VoI to be the absolute first moment of the particle's position $\mathcal{O}(x) = |x|$, we get
\begin{eqnarray*}
 && \left. \langle |x(t)| \rangle \right|_{\lambda + \Delta \lambda} -  \left. \langle |x(t)| \rangle \right|_{\lambda} = \Delta \lambda \left \langle |x (t)| \left( \frac{1}{2 \lambda}-\frac{|x(0)|}{2 \sqrt{\lambda D}} \right) \right \rangle \\
 && = \qquad \frac{\Delta \lambda}{2 \lambda} \left(\langle |x| \rangle - \sqrt{\frac{\lambda}{D}} \langle |x(t)| |x(0)| \rangle \right). 
\end{eqnarray*}

We note that all three examples correspond to the same formal structure, i.e. rely on the same formal FRR. The only difference between these examples is due to the fact that
in (close to) equilibrium cases the corresponding PDFs follow immediately from the postulates of statistical mechanics while in the NESS
case the corresponding PDF had to be calculated explicitly (in Ref.~ \cite{EvaMaj}) before the formal construct gets useful. 

\section{Formal FRRs beyond stationarity \label{sec:Main}}

The existence of a stationary state is a necessary condition for applying the approach of Sec.~\ref{sec:Stand}, but the approach can be generalized to the cases when
we don't start from the stationary state, or the stationary state is absent outright. In this case, we choose the changes in forcing to manipulate not the initial stationary state
 but the transition PDF.

To do so, we consider the following situation. The system is prepared at $t_0$ in a given state characterized by the PDF $p(\mathbf{x}_0|t_0)$.
Let $p(\mathbf{x},t | \mathbf{x}', t' ; F)$ with $t_0 \leq t' < t$ be the transition PDF from the state $\mathbf{x}'$ at time $t'$ to the state $\mathbf{x}$ at time $t$
under constant forcing $F$. To shorten the notation in the following lengthy formulas, we will omit $t_0$ in the condition, and time arguments in probability densities: $p(\mathbf{x}_1 | \mathbf{x}_2 ; F) = p(\mathbf{x}_1,t_1 | \mathbf{x}_2, t_2 ; F)$,
$p(\mathbf{x}_0) = p(\mathbf{x}_0,t_0)$. 
For the observable $\mathcal{O}$ we then have
\begin{equation}
\! \!\! \! \!\langle \mathcal{O}(t) | F_- \rangle =  \int \! \!\! \! \int \!\! \! \!\int    d \mathbf{X} \mathcal{O}(\mathbf{x})  p(\mathbf{x} | \mathbf{x}'; 0) p(\mathbf{x}' | \mathbf{x}_0 ; F)  p(\mathbf{x}_0)
 \label{eq:Fminus1}
\end{equation}
and
\begin{equation}
\! \!\! \! \langle \mathcal{O}(t) | F_+ \rangle =  \int \! \!\! \! \int \!\! \! \!\int   d \mathbf{X} \mathcal{O}(\mathbf{x})  p(\mathbf{x} | \mathbf{x}' ; F) p(\mathbf{x}' | \mathbf{x}_0 ; 0)  p(\mathbf{x}_0)
 \label{eq:Fplus1}
\end{equation}
with $d \mathbf{X}  = d \mathbf{x} d \mathbf{x}' d \mathbf{x}_0$, in the $F_-$ and $F_+$ setup, respectively.
We note that Eqs.~(\ref{eq:Fminus1}) and (\ref{eq:Fplus1}) appear by substituting Eq.~(\ref{eq:CK0}) (under the corresponding choice of transition probabilities for the force on / off) into Eq.~(\ref{eq:CInt}), and then averaging $\mathcal{O}(\mathbf{x})$ over the corresponding distribution.  
We will also need the formula for the case when the force is switched on at $t_0 + \epsilon$ and never switched off:
\begin{eqnarray}
&& \langle \mathcal{O}(t) | F \rangle = \int \! \!\! \! \int   d \mathbf{x} d \mathbf{x}_0 \mathcal{O}(\mathbf{x})  p(\mathbf{x} | \mathbf{x}_0; F) p(\mathbf{x}_0) \label{eq:Fconst} \\
 &&= \int \! \!\! \! \int \!\! \! \!\int   d \mathbf{x} d \mathbf{x}_1 d \mathbf{x}_0 \mathcal{O}(\mathbf{x})  p(\mathbf{x} | \mathbf{x}_1; F) p(\mathbf{x}_1 | \mathbf{x}_0; F)  p(\mathbf{x}_0)
\nonumber  
\end{eqnarray}
with an arbitrary break point at $t_1 \in (t_0,t)$. Note that the second line of Eq.~(\ref{eq:Fconst}) immediately follows from Eq.~(\ref{eq:CK0}), and essentially states that 
the situation under constant force is equivalent to the one the when the force $F$ was switched on immediately after preparation, 
then switched off at some $t_0< t' < t$, and immediately switched on again. 

Let us first turn to the $F_+$ setup, and put down
\begin{eqnarray*}
&& \langle \mathcal{O}(t) | t_0, F_+ \rangle - \langle \mathcal{O}(t) | t_0, 0 \rangle \\ 
&& = \int \! \!\! \! \int \!\! \! \!\int   d \mathbf{x} d \mathbf{x}' d \mathbf{x_0} \mathcal{O}(\mathbf{x})  p(\mathbf{x} | \mathbf{x}'; F) p(\mathbf{x}'| \mathbf{x}_0; 0)  p(\mathbf{x}_0) \\
&& \; \; -\int \! \!\! \! \int \!\! \! \!\int   d \mathbf{x} d \mathbf{x}' d \mathbf{x_0} \mathcal{O}(\mathbf{x})  p(\mathbf{x} | \mathbf{x}'; 0) p(\mathbf{x}' | \mathbf{x}_0; 0)  p(\mathbf{x}_0).
\end{eqnarray*}
Introducing 
\begin{equation}
\widetilde{\mathcal{Y}}(\mathbf{x},t | \mathbf{x}', t' ; F_+) = \frac{p(\mathbf{x},t | \mathbf{x}', t' ; F_+) - p(\mathbf{x},t | \mathbf{x}', t' ; 0)}{p(\mathbf{x},t | \mathbf{x}', t' ; 0)}
\label{eq:Ytilde}
\end{equation}
we may write:
\begin{eqnarray*}
 && \langle \mathcal{O}(t) | t_0, F_+ \rangle - \langle \mathcal{O}(t) | t_0, 0 \rangle = \int \! \!\! \! \int \!\! \! \!\int   d \mathbf{x} d \mathbf{x}' d \mathbf{x_0} \mathcal{O}(\mathbf{x})  \times \\
 &&  \; \;  \widetilde{\mathcal{Y}}(\mathbf{x},t | \mathbf{x}', t' ; F_+)  p(\mathbf{x},t | \mathbf{x}', t' ; 0) p(\mathbf{x}',t' | \mathbf{x}_0, t_0 ; 0)  p(\mathbf{x}_0,t_0) \\
&& \equiv \left\langle \mathcal{O}(\mathbf{x})  \widetilde{\mathcal{Y}}(\mathbf{x},t | \mathbf{x}', t' ; F) \right\rangle_0, 
\end{eqnarray*}
where we have restored temporal arguments for the time being. This is the general form of a non-liner FRR in the $F_+$ setup. 
The average in the last line is taken over the tree-point joint PDF of relevant variables in the absence of the forcing:
\begin{eqnarray}
&& \left\langle \mathcal{O}(\mathbf{x})  \widetilde{\mathcal{Y}}(\mathbf{x},t | \mathbf{x}', t' ; F) \right\rangle_0 = \label{eq:ThreePoint}\\
&&  \; \; \int \! \!\! \! \int  \!\! \! \!\int  \mathcal{O}(\mathbf{x}) \widetilde{\mathcal{Y}}(\mathbf{x},t | \mathbf{x}', t' ; F_+)  p(\mathbf{x},t; \mathbf{x}',t' ; \mathbf{x}_0,t_0) d \mathbf{x} d \mathbf{x}' d \mathbf{x_0} . \nonumber
\end{eqnarray}
In the linear response domain we assume $p(\mathbf{x},t | \mathbf{x}', t' ; F)$ to be differentiable w.r.t. $F$ and approximate the numerator in Eq.~(\ref{eq:Ytilde}) via the derivative. Then
$\widetilde{\mathcal{Y}}(\mathbf{x},t | \mathbf{x}', t' ; F) = F \mathcal{Y}(\mathbf{x},t | \mathbf{x}', t')$
with
\begin{equation}
\mathcal{Y}(\mathbf{x},t | \mathbf{x}', t') = \nabla_{F} \left. \ln p(\mathbf{x},t | \mathbf{x}', t' ; F) \right|_{F=0}. 
 \label{eq:DynConj}
\end{equation}
Thus, we get a formal FRR in the $F_+$ setup:
\begin{equation}
 \langle \mathcal{O}(t) |t_0; F_+ \rangle - \langle \mathcal{O}(t) | t_0; 0 \rangle = F \left\langle \mathcal{O}(\mathbf{x}) \mathcal{Y}(\mathbf{x},t | \mathbf{x}', t') \right\rangle_0.
 \label{eq:FplusF1}
\end{equation}
The dynamics of any single-time VoI is defined by the \textit{dynamic conjugate} to the force, $\mathcal{Y}(...)$, which
explicitly depends on the values of state variables at two times. The mean on the r.h.s. has the same meaning and structure as the one in Eq.~(\ref{eq:ThreePoint}) with the only difference that
$\widetilde{\mathcal{Y}}(\mathbf{x},t | \mathbf{x}', t' ; F)$ is changed for $F \mathcal{Y}(\mathbf{x},t | \mathbf{x}', t')$. 
The discussion of the $F_-$ setup follows along the same lines, and leads to 
\begin{equation}
 \langle \mathcal{O}(t) | t_0; F_- \rangle - \langle \mathcal{O}(t) | t_0; 0 \rangle = F \left\langle \mathcal{O}(\mathbf{x}) \mathcal{Y}(\mathbf{x'},t' | \mathbf{x}_0, t_0) \right\rangle_0.
 \label{eq:FminusF1}
\end{equation}
The formal FRRs, Eqs.~(\ref{eq:FplusF1}) and (\ref{eq:FminusF1}), are the main results of the present work. We note that the only assumption invoked except for Markovianity is 
the differentiability of the transition PDF w.r.t. constant applied force. 

\subsection{The response functions}

The response functions following from Eqs.~(\ref{eq:FplusF1}) and (\ref{eq:FminusF1}) by using Eq.~(\ref{eq:2t}) are given by
\begin{eqnarray}
 \chi_+(t,t',t_0) &=& - \frac{d}{dt'} \left\langle \mathcal{O}(\mathbf{x}) \mathcal{Y}(\mathbf{x},t | \mathbf{x}', t') \right\rangle_0 , \label{eq:chiplus} \\
\chi_-(t,t',t_0) &=& \frac{d}{dt'} \left\langle \mathcal{O}(\mathbf{x}) \mathcal{Y}(\mathbf{x}',t' | \mathbf{x}_0, t_0) \right\rangle_0 .  \label{eq:chiminus}
\end{eqnarray}
Eqs.~(\ref{eq:chiplus}) and (\ref{eq:chiminus}) lead to the same response function $\chi_+(t,t',t_0)=\chi_-(t,t',t_0)=\chi(t,t',t_0)$. 
To show this we consider the situation when the force $F$ was switched on immediately after preparation and never switched off, which, as we already discussed,
is equivalent to the case when the force $F$ was switched on immediately after preparation, 
then switched off at $t'$, and immediately switched on again. This is exactly the $F_+$ case with $t' = t_0$, and in this case
\begin{equation}
  \langle \mathcal{O}(t) | F \rangle - \langle \mathcal{O}(t) | 0 \rangle = F \left\langle \mathcal{O}(\mathbf{x}) \mathcal{Y}(\mathbf{x},t | \mathbf{x}_0, t_0) \right\rangle_0.
  \label{eq:Oneside}
\end{equation}
This expression immediately follows from the first line of Eq.~(\ref{eq:Fconst}):
\begin{eqnarray*}
&& \langle \mathcal{O}(t) | F \rangle - \langle \mathcal{O}(t) | 0 \rangle \\
&&= \int \! \!\! \! \int   d \mathbf{x} d \mathbf{x}_0 \mathcal{O}(\mathbf{x})  [p(\mathbf{x} | \mathbf{x}_0; F) - p(\mathbf{x} | \mathbf{x}_0; 0)] p(\mathbf{x}_0)  \\
&& \simeq F \int \! \!\! \! \int   d \mathbf{x} d \mathbf{x}_0 \mathcal{O}(\mathbf{x})  \left. \nabla_F p(\mathbf{x} | \mathbf{x}_0; F) \right|_{F=0} p(\mathbf{x}_0) \\
&& \equiv F \int \! \!\! \! \int   d \mathbf{x} d \mathbf{x}_0 \mathcal{O}(\mathbf{x})  \left. \nabla_F \ln p(\mathbf{x} | \mathbf{x}_0; F) \right|_{F=0} p(\mathbf{x} | \mathbf{x}_0; 0) p(\mathbf{x}_0) \\
&& = F \left\langle \mathcal{O}(\mathbf{x}) \mathcal{Y}(\mathbf{x},t | \mathbf{x}_0, t_0) \right\rangle_0.
\end{eqnarray*}
Now we turn to the second line of the same Eq.~(\ref{eq:Fconst}): We get
\begin{eqnarray}
&& \langle \mathcal{O}(t) | F \rangle - \langle \mathcal{O}(t) | 0 \rangle \nonumber \\
 && = \int \! \!\! \! \int \!\! \! \!\int   d \mathbf{X} \mathcal{O}(\mathbf{x}) p(\mathbf{x}_0) \times \nonumber  \\
 && \qquad [ p(\mathbf{x} | \mathbf{x}_1; F) p(\mathbf{x}_1 | \mathbf{x}_0; F)- p(\mathbf{x} | \mathbf{x}_1; 0) p(\mathbf{x}_1 | \mathbf{x}_0; 0) ] \label{eq:int} \\
 && \simeq F \int \! \!\! \! \int  \!\! \! \!\int  d \mathbf{X} \mathcal{O}(\mathbf{x}) p(\mathbf{x}_0)
\left. \nabla_F [ p(\mathbf{x} | \mathbf{x}_1; F) p(\mathbf{x}_1 | \mathbf{x}_0; F)] \right|_{F=0}  \nonumber
\end{eqnarray}
with $d \mathbf{X} = d \mathbf{x} d \mathbf{x}_1 d \mathbf{x}_0$. Now we use the differential identity 
\begin{eqnarray*}
&& \nabla_{F} [p(\mathbf{x} | \mathbf{x}_1 ; F)  p(\mathbf{x}_1 | \mathbf{x}_0 ; F)] = 
[\nabla_{F} p(\mathbf{x} | \mathbf{x}_1 ; F)] p(\mathbf{x}_1 | \mathbf{x}_0 ; F)  \\
&& \qquad + p(\mathbf{x} | \mathbf{x}_1 ; F) [\nabla_F p(\mathbf{x}_1 | \mathbf{x}_0 ; F)]
\end{eqnarray*}
applying it at $F=0$. Using the definition of $\mathcal{Y}$, e.g.,
\[
  \mathcal{Y}(\mathbf{x},t | \mathbf{x}_1, t_1) = \frac{\left. \nabla_{F} p(\mathbf{x} | \mathbf{x}_1 ; F) \right|_{F=0}}{p(\mathbf{x} | \mathbf{x}_1 ; 0)} 
\]
we may rewrite the last expression in Eq.~(\ref{eq:int}) as
\begin{eqnarray*}
&& \int \! \!\! \! \int  \!\! \! \!\int \mathcal{O}(\mathbf{x}) p(\mathbf{x}_0)
\left. \nabla_F [ p(\mathbf{x} | \mathbf{x}_1; F) p(\mathbf{x}_1 | \mathbf{x}_0; F)] \right|_{F=0} \\
&& = \int \! \!\! \! \int  \!\! \! \!\int   d \mathbf{X}  \mathcal{O}(\mathbf{x}) p(\mathbf{x}_0) \mathcal{Y}(\mathbf{x},t | \mathbf{x}_1, t_1) p(\mathbf{x} | \mathbf{x}_1; 0) p(\mathbf{x}_1 | \mathbf{x}_0; 0) \\
&& + \int \! \!\! \! \int  \!\! \! \!\int   d \mathbf{X}  \mathcal{O}(\mathbf{x}) p(\mathbf{x}_0) \mathcal{Y}(\mathbf{x}_1,t_1 | \mathbf{x}_0, t_0) p(\mathbf{x} | \mathbf{x}_1; 0) p(\mathbf{x}_1 | \mathbf{x}_0; 0).
\end{eqnarray*}
Therefore,
\begin{eqnarray}
 \langle \mathcal{O}(t) | F \rangle - \langle \mathcal{O}(t) | 0 \rangle &=& F \left\langle \mathcal{O}(\mathbf{x}) \mathcal{Y}(\mathbf{x},t | \mathbf{x}_1, t_1) \right\rangle_0 \\
 && + F \left\langle \mathcal{O}(\mathbf{x}) \mathcal{Y}(\mathbf{x}_1,t_1 | \mathbf{x}_0, t_0) \right\rangle_0. \nonumber 
 \label{eq:Otheside}
\end{eqnarray}
Equating the r.h.s. of Eqs. (\ref{eq:Oneside}) and (\ref{eq:Otheside}) we get 
\begin{eqnarray}
 && \left\langle \mathcal{O}(\mathbf{x}) \mathcal{Y}(\mathbf{x},t | \mathbf{x}_0, t_0) \right\rangle_0 = \left\langle \mathcal{O}(\mathbf{x}) \mathcal{Y}(\mathbf{x},t | \mathbf{x}', t') \right\rangle_0 \nonumber \\
 && \qquad + \left\langle \mathcal{O}(\mathbf{x}) \mathcal{Y}(\mathbf{x}',t' | \mathbf{x}_0, t_0) \right\rangle_0
 \label{eq:Dec}
\end{eqnarray}
giving a chain rule for correlation functions when splitting the interval $(t_0,t)$ at an intermediate point $t'$.
Taking the derivative of the both parts of Eq.~(\ref{eq:Dec}) w.r.t. $t'$ we get
\[
 0 = \frac{d}{dt'} \left\langle \mathcal{O}(\mathbf{x}) \mathcal{Y}(\mathbf{x},t | \mathbf{x}', t') \right\rangle_0 +  \frac{d}{dt'} \left\langle \mathcal{O}(\mathbf{x}) \mathcal{Y}(\mathbf{x}',t' | \mathbf{x}_0, t_0) \right\rangle_0,
\]
and therefore $\chi_+(t,t',t_0) = \chi_-(t,t',t_0)$. 

Our discussion can be repeated for the case when the force is a vector $\vec{F}$ with $N$ components $F_\alpha$,
and $\mathcal{O}$ a vector with $M$ components $\mathcal{O}_i$. The response function is now a matrix $\chi_{i, \alpha}(t,t',t_0)$, so that 
$ \langle \mathcal{O}_i(t)| t_0; \{\vec{F} \} \rangle - \langle \mathcal{O}_i(t) | t_0; 0 \rangle = \sum_{\alpha=1}^N \int_{t_0}^t \chi_{i, \alpha}(t,t',t_0) F_\alpha(t') dt'$. The dynamic conjugate is a vector
$\vec{\mathcal{Y}}(\mathbf{x},t | \mathbf{x}', t') = \nabla_{\vec{F }} \left. \ln p(\mathbf{x},t | \mathbf{x}', t' ; F) \right|_{\vec{F}=0}$ with the same number of components as $\vec{F}$ 
(since $p(...)$ is a scalar), and the relations Eq.~(\ref{eq:chiplus}) and (\ref{eq:chiminus}) read:
\begin{eqnarray*}
  \chi_{+,i,\alpha}(t,t',t_0) &=& - \frac{d}{dt'} \left\langle \mathcal{O}_i(\mathbf{x}) \mathcal{Y}_\alpha (\mathbf{x},t | \mathbf{x}', t') \right\rangle_0 ,  \\
\chi_{-, i, \alpha}(t,t',t_0) &=& \frac{d}{dt'} \left\langle \mathcal{O}_i(\mathbf{x}) \mathcal{Y}_\alpha(\mathbf{x}',t' | \mathbf{x}_0, t_0) \right\rangle_0 .
\end{eqnarray*}

\subsection{Relation to previous results}

Let us consider the $F_-$ setup and discuss the case when the system at constant forcing tends to a stationary state which does not depend on initial conditions. 
In this case
$\lim_{t_0 \to - \infty} p(\mathbf{x}',t' | \mathbf{x}_0, t_0 ; F) = p(\mathbf{x}'| F)$ with $p(\mathbf{x}'| F)$ being the steady state at $t'$.
Now,
\begin{equation}
\lim_{t_0 \to - \infty}  \mathcal{Y}(\mathbf{x}',t' | \mathbf{x}_0, t_0) = \nabla_{F} \left. \ln p(\mathbf{x}' | F) \right|_{F=0} = Y(\mathbf{x}').
 \label{eq:Dec1}
\end{equation}
The variable $ Y(\mathbf{x}')$ is a usual (i.e., static) conjugate to the force, as discussed in Sec. \ref{sec:Stand}.  
In this case, the mean on the r.h.s. of Eq.~(\ref{eq:FminusF1}) reads:
\begin{eqnarray*}
&& \int \! \!\! \! \int  \!\! \! \!\int  \mathcal{O}(\mathbf{x}) Y(\mathbf{x}')  p(\mathbf{x},t; \mathbf{x}',t' ; \mathbf{x}_0,t_0) d \mathbf{x} d \mathbf{x}' d \mathbf{x_0} \\
&& = \int \! \!\! \! \int  d \mathbf{x} d \mathbf{x}' \mathcal{O}(\mathbf{x}) Y(\mathbf{x}') p(\mathbf{x} | \mathbf{x}'; t') \left[\int d\mathbf{x}_0  p(\mathbf{x}'| \mathbf{x}_0; 0) p(\mathbf{x}_0) \right]  \\
&& = \int \! \!\! \! \int   d \mathbf{x} d \mathbf{x}' \mathcal{O}(\mathbf{x}) Y(\mathbf{x}') p(\mathbf{x}, t;\mathbf{x}'; t') p(\mathbf{x}')
\end{eqnarray*}
where in the last line, for the expression in square brackets, we used the fact that the stationary distribution is invariant under force-free evolution, $\int d\mathbf{x}_0  p(\mathbf{x}'| \mathbf{x}_0; 0) p(\mathbf{x}_0) = p(\mathbf{x}')$.
Thus, in this case, the $F_-$ setup reproduces the result, Eq.~(\ref{eq:FDTStan}), of the standard approach.  The single-time nature of static conjugate makes the $F_-$ setup simpler for stationary cases.

The $F_-$ variant of our FRR is closely related to the modified FDT of Verley et al. \cite{Verley}, where our approach is much more straightforward. 
Ref.~\cite{Verley} does not stress the two-time structure of the conjugate which is however evident from intermediate calculations. To see the relation, let us return to 
Eq.~(\ref{eq:Fminus1}), and note that $p_1(\mathbf{x}', t'; t_0, F_-) = \int d \mathbf{x}_0 p(\mathbf{x}',t' | \mathbf{x}_0,t_0 ; F_-)  p(\mathbf{x}_0,t_0)$ 
is the PDF of $\mathbf{x}$ at time $t'$ when the force was switched off.  
Then we can perform the same steps as before, now keeping $p(\mathbf{x}',t' | \mathbf{x}_0,t_0 ; F_-)  p(\mathbf{x}_0,t_0)$ together, and introduce the new 
conjugate $\mathcal{Z}(\mathbf{x}', t';t_0) = \left. \nabla_F \ln p_1(\mathbf{x}', t'; t_0, F) \right|_{F=0}$. 
Then, $\chi(t,t',t_0) = \frac{d}{dt'} \left\langle \mathcal{O}(\mathbf{x}) \mathcal{Z}(\mathbf{x}', t';t_0)\right\rangle_0$. Note that our $\mathcal{Z}$ 
is nothing else than $-\left. \partial_h \psi(t',c) \right|_{h \to 0}$ of
Ref.~\cite{Verley}.  At difference to a 
universal dynamic conjugate $\mathcal{Y}$, the $\mathcal{Z}$-conjugate does depend on initial conditions, and has to be recalculated when these conditions change.

\subsection{A relation between static and dynamic responses}

An interesting additional result pertinent to systems starting in a stationary state is the integral identity following from the decoupling, Eq.~(\ref{eq:Dec1}), and the chain rule, Eq.~(\ref{eq:Dec}). 
We again note that in the absence of the perturbation a system prepared long ago tends to a stationary state. Sending $t_0$ to $-\infty$ we see that the l.h.s. of Eq.~(\ref{eq:Dec}) and the second term on 
its r.h.s. are calculated taking the values of $\mathcal{Y}$ in stationary state, i.e., essentially $Y$: 
\begin{equation}
\langle \mathcal{O}(t) Y(t) \rangle_0 = \langle \mathcal{O}(t) \mathcal{Y}(t,t') \rangle_0 + \langle \mathcal{O}(t) Y(t') \rangle_0 .
\label{eq:II}
\end{equation}
The values of $Y$ as functions of $\mathbf{x}$ are not explicitly time-dependent, and the mean $\langle \mathcal{O}(t) Y(t') \rangle_0$ can only depend on the time lag $\Delta t = t-t'$.  
Now we interpret the l.h.s. of Eq.~(\ref{eq:II}), where the time lag $\Delta t = 0$. We return to our Eq.~(\ref{eq:FDTStan}), and consider the value 
of the response immediately after switching off the force. We get: $\langle \mathcal{O}(0)| F_- \rangle - \langle \mathcal{O}(0)| 0 \rangle =  F \left\langle  \mathcal{O}(0) Y(0) \right\rangle_0$, so that
the value $\mathcal{X}_{\mathcal{O}F} = \left\langle  \mathcal{O}(0) Y(0) \right\rangle_0 \equiv \left\langle  \mathcal{O}(t) Y(t) \right\rangle_0$ corresponds to the static susceptibility $\mathcal{X}_{\mathcal{O}F}$ of the stationary system to a constant force.
We therefore have:
\begin{equation}
 \langle \mathcal{O}(t) Y(t') \rangle_0 + \langle \mathcal{O}(t) \mathcal{Y}(t,t') \rangle_0 = \mathcal{X}_{\mathcal{O}F}.
\end{equation}
Let us stress that if the initial condition corresponds to a stationary distribution the introduction of the dynamical conjugate might be considered as an unnecessary complication, completing the whole picture but not
bringing computational advantages. The situation is different if we don't start from a stationary distribution, and it changes dramatically if the stationary state is non-existent.

\section{Simple examples}

In this Section we consider several simple examples illustrating important properties of dynamic conjugates.
All our examples, which build on each other,  will correspond to the case similar to the one considered in Sec. \ref{sec:Bilinear}, namely to the action of the constant force, which really constitutes the simplest case.
We will, however, consider several different situations, and see that comparing them is quite instructive.

\subsection{Example 4. Biased Brownian motion}

We start our discussion from the really simplest case of Brownian motion biased by the action of a constant force $F$, as described by the Langevin equation
\[
 \dot{x} = \mu F + \sqrt{2 D} \xi(t),
\]
with the mobility $\mu$ and the diffusion coefficient $D$ connected by the Einstein's relation $D=\mu k_B T$, and  $\xi(t)$ being the standard Gaussian white noise with $\langle \xi(t) \rangle = 0$, $\langle \xi(t) \xi(t)\rangle = \delta(t-t')$.
The conditional PDF of the displacement is 
\[
 p(x,t|x_0,t_0; F) = \frac{1}{\sqrt{4 \pi D (t-t_0)}} e^{- \frac{[x-x_0-\mu F(t-t_0)]^2}{4 D (t-t_0)}}.
\]
Using this form we obtain 
\begin{equation}
 \mathcal{Y}(x,t;x_0,t_0) = \frac{\mu}{2 D}(x-x_0) = \frac{1}{2 k_B T} (x-x_0).
 \label{eq:YdynStand}
\end{equation}
Let us note that our system does not possess the stationary state since  $p(x,t|x_0,t_0)$, keeping its normalization at all finite $t$, tends to zero for any $x$ when $t \to \infty$. 
Therefore, using our non-stationary approach is the only option. 
Let us fix the initial condition  $x_0$ at $t_0=0$, and consider the increment $\Delta x =x-x_0$ as a VoI  (fixing initial conditions makes the VoI a single-time one). 
In the absence of the external force we have $\langle \Delta x|0 \rangle =0$, and the FRR reads:
\begin{equation}
 \langle \Delta x | F \rangle = \frac{F}{2 k_B T} \langle (\Delta x)^2 \rangle.
 \label{eq:GER1}
\end{equation}
Knowing that $ \langle \Delta x | F \rangle = \mu f (t-t_0)$ and $  \langle [\Delta x(t-t_0)]^2 \rangle = 2 D (t-t_0)$, we see that the result is essentially 
the Einstein's relation $\mu = D/(k_B T)$ (which was assumed at the very beginning of our discussion). 

Comparing our dynamic conjugate, Eq.~(\ref{eq:YdynStand}), with the standard case, Eq.~(\ref{eq:Ystand}) we see two differences:
The presence of the initial condition, and the value of the prefactor of the coordinate, being $\beta = (k_B T)^{-1}$ in $Y$, and $\frac{1}{2} \beta$ in $\mathcal{Y}$. 
Equations (\ref{eq:Ystand}) and (\ref{eq:YdynStand}) are the two different limiting cases of the same expression. 
We will return to this discussion after considering the next example. 

\subsection{Example 5. Ornstein-Uhlenbeck process in and out of equilibrium \label{Sec:VelOU}}

Our next simple example corresponds to the Ornstein-Uhlenbeck (OU) process \cite{OU}, namely, we consider a velocity of a massive Brownian particle with mass $m$ and friction coefficient $\gamma = \mu^{-1}$ moving
under the influence of an external force, as described by the Langevin equation
\begin{equation}
 m \dot{v} = - \gamma v + F + \sqrt{2 \gamma k_B T} \xi(t).
 \label{eq:OUVel}
\end{equation}
A combination $\tau = m / \gamma$ defines the relaxation time of velocity, and its inverse will be denoted by $\alpha = 1/\tau = \gamma/m$ in what follows.
For the following discussion we note that the equation is formally the same as the one describing the motion of an overdamped particle in a harmonic potential with the spring strength $k$,
\begin{equation}
\gamma \dot{x} = - k x + F + \sqrt{2 \gamma k_B T} \xi(t).
\label{eq:OUOver}
\end{equation}
This last equation formally ensues by putting the l.h.s. of Eq.~(\ref{eq:OUVel}) to zero (i.e., formally taking $m=0$), and splitting the force into internal harmonic part $-kx$ and external constant one $F$. 
The translation of the results between these two situations is reached by renaming $m \to \gamma$, $\gamma \to k$, $\tau = \gamma/k$.

Now we concentrate on the Eq.~(\ref{eq:OUVel}). Our aim is to show explicit forms for different conjugates, and to see how known results arise from formal FRRs.
The details of calculations are given in Appendix \ref{Sec:OU}. We set $t_0 = 0$, and omit it from the list of arguments of all functions. Using the known form of the transition PDF, Eq.~(\ref{eq:ConDis}), we obtain:
\begin{equation}
  \mathcal{Y}(v,t;v',t') = \frac{\mu m}{k_B T} \frac{v-v' e^{-\alpha(t-t')}}{1+e^{-\alpha(t-t')}},
  \label{eq:ConjVel}
\end{equation}
and see that this is explicitly time-dependent. Now we consider the $F_+$ setup. We get:
\begin{equation}
\langle v(t) | F_+ \rangle - \langle v(t) | 0 \rangle =\frac{\mu m F \left[\langle v^2(t) \rangle - \langle v(t) v(t') \rangle e^{-\alpha(t-t')}  \right]}{k_B T[1+e^{-\alpha(t-t')}]} , \label{eq:FplusEx}
\end{equation}
where the r.h.s. is calculated for given initial conditions but without forcing. When starting from the equilibrium state, we have
$\langle v(t) | 0 \rangle =0$, $\langle v^2(t)\rangle = \frac{k_B T}{m}$,  $\langle v(t) v(t')  \rangle = \frac{k_B T}{m}  e^{-\alpha(t-t')}$,
so that
\[
\langle v(t) | F_+ \rangle - \langle v(t) | 0 \rangle = \mu F \frac{1-e^{-2 \alpha(t-t')}}{1+e^{-\alpha(t-t')}}, 
\]
i.e.,
\[
  \langle v(t) | F_+ \rangle = \mu F [1-e^{-\alpha(t-t')}].
\]  

Starting from a well-prescribed velocity $v_0 \neq 0$ (a non-equilibrium case) we have: $\langle v(t)|0 \rangle = v_0 e^{-\alpha t}$,
$\langle v^2(t) \rangle =\frac{k_B T}{m} (1-e^{-2 \alpha t} + v_0^2 e^{-2 \alpha t}$ and $\langle v(t) v(t') \rangle =
e^{-\alpha (t-t')} \left[k_B T(1-e^{-2 \alpha t'}) + v_0^2 e^{-2 \alpha t'} \right]$.  Substituting the last two expressions into Eq.~(\ref{eq:FplusEx}) we get:
\begin{equation}
\langle v(t) | F_+ \rangle - \langle v(t) | 0 \rangle  = \mu F [1-e^{-\alpha(t-t')} ],
\label{eq:VelRel}
\end{equation}
exactly as in the equilibrium case. Note, however, that the time evolution of the mean velocity in these case is different from the previous one, due to differences in the unperturbed dynamics:
\[
 \langle v(t) | F_+ \rangle =  v_0 e^{-\frac{t}{\mu m}} + \mu F [1-e^{-\alpha(t-t')}].
\]
The response function is given by $\chi(t,t') = m^{-1} e^{-\alpha(t-t')}$, the same as given by the standard FRR. 
The fact of independence of the response function on initial conditions is a peculiarity of the model with exponential relaxation, and is not carried over to more complex situations.

For the further, we note that for fixed $v_0$ we can consider the velocity increment $\Delta v(t) = v(t) - v_0$ for which Eq.~(\ref{eq:VelRel}) holds as well:
\begin{equation}
\langle \Delta v(t) | F_+ \rangle - \langle \Delta v(t) | 0 \rangle  = \mu F [1-e^{-\alpha(t-t')} ],
\label{eq:VelInc}
\end{equation}
since the contributions of $v_0$ cancel on the l.h.s. and $v_0$ is absent on the r.h.s. We will return to this representation in Sec. \ref{Sec:AugOU}.

For the sake of completeness, we also give expressions
for other relevant conjugates, see Appendix \ref{Sec:OU}. Thus, $Y(v) =\frac{\mu m}{k_B T} v$. The $\mathcal{Z}$-conjugates (the ones used in Ref. \cite{Verley}) are:
$\mathcal{Z}(v',t';0) = \frac{\mu m}{k_B T}(1- e^{-t'/\tau}) v'$ when starting at equilibrium, and 
$\mathcal{Z}(v',t';0) = \frac{\mu m}{k_B T} \frac{e^{t'/\tau} v' - v_0}{1+e^{t'/\tau}}$ when starting with velocity $v_0$. Note again that at difference to a 
universal dynamic conjugate $\mathcal{Y}$ these $\mathcal{Z}$-conjugates do depend on initial conditions. 

\subsubsection{Limiting cases for a particle in a harmonic potential}

Before going further with our discussion, let us return to our Eqs.~(\ref{eq:Ystand}) and (\ref{eq:YdynStand}). Using our dictionary to translate between the dynamics of the velocity, and the overdamped case,
we may write for the last case
\[
  \mathcal{Y}(x,t;x_0,t_0) = \frac{1}{k_B T} \frac{x-x_0 e^{-\frac{t-t_0}{\tau} }}{1+e^{-\frac{t-t_0}{\tau} }}.
\]
The equilibrium situation corresponds to sending $t_0 \to - \infty$, so that 
\[
 \lim_{t_0 \to - \infty} \mathcal{Y}(x,t;x_0,t_0) = Y(x) = \frac{x}{k_B T},
\]
Eq.~(\ref{eq:Ystand}).
The situation of the free biased Brownian motion (i.e. the motion in the absence of the harmonic potential) corresponds to taking $k \to 0$, i.e. $\tau \to \infty$,
in which case we get 
\[
 \lim_{\tau \to \infty} \mathcal{Y}(x,t;x_0,t_0) = \frac{1}{k_B T} \frac{x-x_0}{2},
\]
Eq.~(\ref{eq:YdynStand}).

\bigskip

As we have already mentioned, when starting from the stationary distribution, the introduction of the dynamical conjugate might seem to be an unnecessary complication. 
The situation changes dramatically if the stationary state is non-existent like in our first example. Our next example of such a case is even more interesting.

\subsection{Example 6. A non-proper case: A particle on an interval with absorbing boundaries. \label{sec:Nonprop}}

Let us start from stressing that in our discussion we never requested the densities $p(...; {F })$ to be proper PDFs normalized to unity, if only these densities can be used to
obtain the means of $\mathcal{O}(x)$ defined correspondingly. This allows to apply the formalism to improper cases (absorbing states, sources, or birth-death processes), cf. \cite{Padmanabha}. Here, again, we consider only the simplest example.
We note that Ref.~\cite{Padmanabha} used a class B approach, and essentially concentrated only on the long-time behavior of corresponding processes, which were more complex than the one
we consider here.

We consider a particle in a one-dimensional interval $0 \leq x \leq l$ with absorbing boundaries at $0$ and at $l$ under the action
of the additional force.
The non-proper transition PDF by from $x_0$ to $x$ during the time $t$ will be denoted by $w(x|x_0,t)$ (the letter $p$ stays reserved for normalized, proper densities), 
and the initial condition corresponds to $w(x|x_0 ,0) = \delta(x-x_0)$ with $x_0$ being the initial particle's position. 
The density $w(x|x_0,t)$ can be found by solving the 
diffusion equation in presence of the constant force $F$, which reads:
\begin{equation}
\frac{\partial w}{\partial t} = - \mu F \frac{\partial w}{\partial x} +D \frac{\partial^2 w}{\partial x^2} 
\label{eq:Drift} 
\end{equation}
with $D$ again being the diffusion coefficient and $\mu$ being the mobility of the particle. The boundary conditions are: $w(0|x_0,t) = w(l|x_0,t)=0$.

The solution follows from a general discussion given in \cite{Polyanin}, see Sec. 1.1.4.
The way to the solution is given in Appendix \ref{App:Nonprop}. The solution reads:
\begin{eqnarray}
w(x,t|x_0,0;F) && = \exp \left(- \frac{\mu^2 F^2}{4 D} t + \frac{\mu F}{2 D} x \right) \label{eq:Nonp1} \\
 && \qquad \times \exp \left( - \frac{\mu F}{2 D} x_0 \right) G(x,x_0,t) \nonumber
\end{eqnarray}
with $G(x,x_0,t)$ being the Green's function of the diffusion equation on an interval with two absorbing boundaries in the absence of the force,
\[
 G(x,x_0,t) = \frac{2}{l}\sum_{n=1}^\infty \sin\left(\frac{n \pi x}{l}\right) \sin\left(\frac{n \pi x_0}{l}\right) e^{-\frac{D n^2 \pi^2 t}{l^2}}.
\]
Let us concentrate on $F_+$ setup, set $t_0 = 0$, and consider
\begin{eqnarray*}
 \mathcal{Y}(x,t|x_0,0) &=& \left. \frac{\partial}{\partial F} \ln w(x,t|x_0,0;F) \right|_{F=0} \\
 &=& \left. \left[-\frac{\mu^2 F t}{2D} +\frac{\mu}{2D} (x-x_0)\right] \right|_{F=0} \\
 &=& \frac{\mu}{2D}(x-x_0), 
\end{eqnarray*}
which is the same as the one for free Brownian motion, as given by Eq.~(\ref{eq:YdynStand}). At this stage, we don't know whether the form of $\mathcal{Y}(x,t|x_0,0)$ is
generally independent on the boundary conditions, but see that it is the same for free and absorbing ones. 

Now we can use our setup to calculate the response of the simplest observable $\mathcal{O}(x) = 1$ to the external forcing.
Note that the mean value of $\mathcal{O}(x)$ is the survival probability $\langle \mathcal{O}(t) \rangle = \int_0^l w(x|x_0,t) dx = \Phi(t)$, so that essentially we discuss its 
(and therefore the mean first passage time's) response to the force. The change in the survival probability is then related to the force via
\[
 \left. \Phi(t) \right|_{F_-} - \left. \Phi(t) \right|_{F = 0} = \frac{F}{2 k_B T} \langle x - x_0 \rangle,
\]
where the generalized mean
\[
 \langle x - x_0 \rangle = \int_0^l (x - x_0) G(x,x_0,t) dx 
\]
cannot, however, be interpreted as a mean particle's displacement, because the corresponding integral is calculated for a non-proper distribution.

Now we can note that one can normalize $G$ by taking $H(x,x_0,t) = G(x, x_0,t)/\Phi(t)$ being a proper
PDF of the particle's position within the interval conditioned on the survival of the particle. Then 
\[
  \langle x - x_0 \rangle =  \langle x - x_0 \rangle_S \Phi(t),  
\]
where $\langle x - x_0 \rangle_S$ is the true mean position of the survived particles in the absence of the force. Now we get the following FRR:
\[
 \left. \Phi(t) \right|_{F_-} - \left. \Phi(t) \right|_{F = 0} = \frac{F}{2 k_B T} \langle x - x_0 \rangle_S \left. \Phi(t) \right|_{F = 0}.
\]
One can immediately infer that the 
linear response to the force vanishes, if the particle starts at the middle of the interval, due to symmetry. If the particle starts close to the left boundary,
application of the force showing to the right leads to increasing the survival probability (simply because it moves the particle further from the closest ``dangerous'' absorbing point
making the survival more probable), etc. 

\subsection{Example 7. Underdamped Brownian particle: A ``non-minimal'' situation \label{Sec:AugOU}}

Up to now, all our examples corresponded to systems whose Markovian description only needed a single variable (i.e., the ones with a one-dimensional state space),
and VoIs depending on this variable only.
Now we consider a slightly more complicated system corresponding to the Brownian motion of a massive particle, with the Langevin equation
in the form
\[
 m \ddot{x} = F - \gamma \dot{x}(t) + \sqrt{2 k_B T \gamma} \xi(t)
\]
i.e. with the equations for the state vector  $\mathbf{s} = (x,v)^{\mathrm{T}}$
\begin{eqnarray}
\dot{x} &=& v \label{eq:x} \\
 m\dot{v} &=& F - \gamma v(t) + \sqrt{2 k_B T \gamma} \xi(t).\label{eq:v} 
\end{eqnarray}
Note that Eq.~(\ref{eq:v}) is autonomous (it is the same as the previous Eq.~(\ref{eq:OUVel}) ), so that the minimal list consists of a single variable $v$. We are, however, interested in displacements $\Delta x$, and have to 
use the augmented list. This example is by far not as trivial as it might seem.
´

The problem under constant force switched on at $t=0$ can be reduced to a problem without the
force by changing into a reference frame moving with velocity $V = F/\gamma$, i.e. performing a Galilean transformation. This changes only the means (i.e. centerings) but not the central moments
of the displacements. Introducing the new velocity $u = v(t) - F/\gamma$, and the new coordinate $y = x - Ft/\gamma$
reduces the equations to the ones without forcing:
\begin{eqnarray*}
\dot{y} &=& u\\
 m\dot{u} &=& - \gamma u(t) + \sqrt{2 k_B T \gamma} \xi(t) .
\end{eqnarray*}

The solution for PDF for the corresponding initial condition problem was given by S. Chandrasekhar in 1943 \cite{Chandra}. In what follows, we will introduce the 
``force'' $f= F/\gamma$ (with the dimension of velocity $[f] = \mathrm{\frac{L}{T}}$, being equal to the drift velocity of a particle in the force
field in the absence of the noise) and $\alpha = 1/\tau = \gamma / m$ being the inverse relaxation time as before. 

The initial condition for $y$ is the same as for $x$, but the initial condition for $u$ is now different from the one for $v$, since 
\[
 u(0) = v(0) -  f.
\]
The solution for the PDF in variables $(y,u)$ reads

\begin{eqnarray*}
 p(y,u;t) = &&\frac{1}{2\pi \sigma_x \sigma_v \sqrt{1-\rho^2}}\exp \left[-\frac{1}{2(1-\rho^2)} \left(\frac{(y-\mu_y)^2}{\sigma_x^2} \right. \right. \\
 && \left. \left.-\frac{2 \rho(y-\mu_y)(u-\mu_u)}{\sigma_x\sigma_v}+\frac{(u-\mu_u)^2}{\sigma_v^2(t)} \right)\right],
\end{eqnarray*}
with parameters $\mu_u(t), \mu_y(t)$ which depend on the initial conditions and on the force, 
\[
 \mu_u(t) =  e^{-\alpha t} v_0 - e^{-\alpha t} f, 
 \]
 \[
  \mu_y(t) = x_0 + \frac{1-e^{-\alpha t}}{\alpha} v_0 - \frac{1-e^{-\alpha t}}{\alpha} f ,
 \]
and $\sigma_v(t), \sigma^2_x(t)$ and $\rho(t)$, which are independent of them:

\begin{eqnarray}
 \sigma^2_v(t) &=& v_{eq}^2 (1-e^{-2 \alpha t}) , \nonumber \\
 \sigma^2_x(t) &=& v_{eq}^2 \frac{2\alpha t-3+4e^{-\alpha t}-e^{-2\alpha t}}{\alpha^2} , \label{eq:SecondMom}\\
 \rho(t) &=& \frac{\langle x(t) v(t) \rangle}{\sigma_x(t)  \sigma_v(t)}= \frac{(1-e^{-\alpha t})^2}{\sigma_x(t)  \sigma_v(t)} \nonumber \\
 &=& \frac{(1-e^{-\alpha t})^2}{\sqrt{(1-e^{-2 \alpha t})(2\alpha t-3+4e^{-\alpha t}-e^{-2\alpha t}})} \nonumber 
\end{eqnarray}
with $v_{eq}^2 = kT/m$.

To investigate the influence of the force, we first can put $x_0 = 0$ (since the $x$-dependence, due to the spatial homogeneity, is anyhow a dependence on
$x-x_0$ only, which will be later denoted as $\Delta x$), but keep the dependence on $v_0$.
The means for $v$ and $x$ are: 
\begin{eqnarray*}
 \mu_v(t) &=& \mu_u(t) + f = e^{-\alpha t} v_0 - e^{-\alpha t} f + f = v_0 e^{-\alpha t} + q(t) f,\\
\mu_x(t)  &=& x_0 + \frac{1-e^{-\alpha t}}{\alpha} v_0 - \frac{1-e^{-\alpha t}}{\alpha} f + ft \\
&=& x_0 + \frac{q(t)}{\alpha} v_0 + p(t) f 
\end{eqnarray*}
with
\begin{equation}
 p(t) = t - \frac{1-e^{-\alpha t}}{\alpha}, \qquad q(t) = 1 - e^{-\alpha t}, \label{eq:AuxFunc}
\end{equation}
while the central second moments stay the same.

The transition probability $p(\mathbf{s}_2,t_2|\mathbf{s}_1,t_1;F_+)$ depends only on $t = t_2 - t_1$, and is a bivariate Gaussian defined by the values of the first and the second
moments:
\begin{widetext}
\begin{eqnarray*}
 p(x,v;t|v_0) = &&\frac{1}{2\pi \sigma_x \sigma_v \sqrt{1-\rho^2}}\exp \left[-\frac{1}{2(1-\rho^2)} \left(\frac{(x-\frac{q(t)}{\alpha} v_0-f p(t))^2}{\sigma_x^2} \right. \right. \\
 && \left. \left.-\frac{2 \rho(x-\frac{q(t)}{\alpha} v_0-f p(t))(v- v_0 e^{-\alpha t} - f q(t))}{\sigma_x\sigma_v}+\frac{(v- v_0 e^{-\alpha t} - f q(t))^2}{\sigma_v^2(t)} \right)\right].
\end{eqnarray*}
Building a marginal PDF for $v$ only, we return to our previous result, Sec. \ref{Sec:VelOU}.

A conjugate to a scalar force $f$ with dimension $[L/T]$ (i.e. of velocity) is a scalar  $ \mathcal{Y}_f(x,v,t|v_0)$ with dimension $T/L$. It is given by 
\[
 Z = \left. \frac{\partial}{\partial f} \ln  p(x,v;t) \right|_{f=0} ,
\]
and is equal to 
\[
 \mathcal{Y}_f(x,v,t|v_0) = \frac{1}{(1-\rho^2)} \left[x \left(\frac{p}{\sigma_x^2} - \frac{\rho q }{\sigma_x \sigma_y} \right) + v \left(\frac{q}{\sigma_v^2} - \frac{p \rho }{\sigma_x \sigma_y} \right) + v_0 \left(\frac{q^2 \rho}{\alpha \sigma_v \sigma_x} + \frac{p e^{-\alpha t} \rho}{\sigma_v \sigma_x} - \frac{p q \rho}{\alpha \sigma_x^2 } - \frac{q e^{-\alpha t}}{\sigma_v^2} \right) \right].
\]
\end{widetext}
Substituting the expressions for the second moments and the correlation function, Eq.~(\ref{eq:SecondMom}), and for the auxiliary functions $p$ and $q$, and simplifying the ensuing expression
(which is lengthy, but \textsc{Mathematica} does the job) we arrive at an astonishingly simple expression,
\[
 \mathcal{Y}_f(x,v,t|v_0) = \frac{v-v_0+\alpha x}{2 v_{eq}^2}
\]
which now does not bear explicit time dependence. Now we can consider a conjugate to the physical force $F= \gamma f$, substitute the values of $\alpha = \gamma/m$ and 
$v_{eq}^2 = k_B T/m$, and get:
\begin{equation}
  \mathcal{Y}(x,v,t|v_0) = \frac{m}{\gamma} \frac{1}{2 k_B T} (v-v_0) + \frac{1}{2 k_B T} (x-x_0),
  \label{eq:YdynKK}
\end{equation}
where $v$ and $x$ are the velocity and position at time $t$, and where we have reintroduced the initial position $x_0$ into the final expression. 
$\mathcal{Y}(x,v,t|v_0)$ has the dimension of the inverse force.

\paragraph{Overdamped limit.}  In the overdamped limit $m \to 0$ we get
\[
 \mathcal{Y}(x) = \frac{1}{2 k_B T} (x-x_0),
\]
proportional to the displacement from the initial position $\Delta x = x-x_0$,  c.f. Eq.~(\ref{eq:YdynStand}). 

\paragraph{Underdamped case.} To further elucidate the result, Eq.~(\ref{eq:YdynKK}), let us start from the sharply defined velocity $v_0$,
still keeping $\Delta x$ as a VoI and taking $t' = 0$.
Using our dynamic conjugate, Eq.~(\ref{eq:YdynKK}), we get:
\begin{equation}
 \langle \Delta x | F \rangle - \langle \Delta x | 0 \rangle =  \frac{F}{2 k_B T} \left( \frac{1}{\alpha} \langle  \Delta x(t) \Delta v(t) \rangle + \langle \Delta x^2(t) \rangle \right)
 \label{eq:FRRKK}
\end{equation}
(with $\Delta v = v-v_0$) where the means on the r.h.s. are taken in the absence of the force. This expression is a generalization of Eq.~(\ref{eq:GER1}) to the
underdamped case, and differs from it by the presence of the first term on the r.h.s. depending on $\langle  \Delta x(t) \Delta v(t) \rangle$. In the limit of fast velocity relaxation, $\alpha \to \infty$, it indeed 
reduces to Eq.~(\ref{eq:GER1}).

To check the relation, we note that from Eq.~(\ref{eq:VelRel}) we have:
\[
 \langle v(t) | F_+ \rangle - \langle v(t) | 0\rangle = \frac{F}{\gamma} (1-e^{-\alpha t})
\]
with 
\[
 \langle v(t) | 0\rangle = v_0 e^{-\alpha t}.
\]
Integrating over $t$ we get our result 
\[
 \langle \Delta x | F \rangle = \frac{q(t)}{\alpha} v_0 + p(t) f = \frac{1-e^{-\alpha t}}{\alpha} v_0 + \frac{F}{\gamma} \left[t-\frac{1-e^{-\alpha t}}{\alpha} \right],
\]
which corresponds to 
\[
 \langle \Delta x | F \rangle - \langle \Delta x | 0 \rangle = \frac{F}{\gamma} \left[t-\frac{1-e^{-\alpha t}}{\alpha} \right].
\]
The explicit calculation of the r.h.s. is tedious, and leads to expressions which have to be rearranged and simplified by hand. We refrain from reproducing these calculations.
The alternative, much simpler, way to calculate the r.h.s. of Eq.~(\ref{eq:FRRKK}) in given in Appendix \ref{Sec:AppUnder}.
The final result is 
\[
  \frac{1}{\alpha} \langle  \Delta x \Delta v \rangle + \langle \Delta x^2 \rangle   = \frac{2 k_B T}{\gamma} \left[t-\frac{1-e^{-\alpha t}}{\alpha} \right],             
\]
showing that the corresponding FRR holds. 

To reproduce our previous result for the velocity only, Eq.~(\ref{eq:VelRel}), we note that this would correspond to 
\[
 \langle  v(t) | F_+ \rangle - \langle v(t) | 0 \rangle  =  \frac{F}{2 k_B T} \left\langle v(t) \left[\frac{1}{\alpha}  \Delta v(t) + \Delta x(t) \right]\right\rangle.
\]
Here, $\langle v(t) \Delta v (t) \rangle = \langle v^2(t) \rangle -v_0 \langle v(t) \rangle$ (both were already calculated in Sec.~\ref{Sec:VelOU}), 
while $\langle  v(t) \Delta x(t) \rangle$ is calculated in App. \ref{Sec:AppUnder}, see Eq.~(\ref{eq:Deltaxv}). Substituting the corresponding 
expressions and simplifying the final result we get
\begin{eqnarray*}
&& \frac{F}{2 k_B T} \left\langle v(t) \left[\frac{1}{\alpha}  \Delta v(t) + \Delta x(t) \right]\right\rangle =\frac{F}{2 k_B T} \frac{2 v_{eq}^2 (1-e^{-\alpha t})}{\alpha} \\
&& =  \mu F (1-e^{-\alpha t}),
\end{eqnarray*}
equivalent to Eq.~(\ref{eq:VelRel}) since now we put $t'=0$. 

Note that while the Eq.~(\ref{eq:YdynKK}) easily reproduces the overdamped limit, the situation with the VoI depending on 
velocities only,  Eq.~(\ref{eq:ConjVel}),  cannot be reproduced trivially by a limiting transition.
The conjugate in the augmented state space, Eq.~(\ref{eq:YdynKK}), does not include the time explicitly, but the effective time
dependence is absorbed into the additional $x$-variable. The equations for the means following from both approaches, however, coincide. 
Therefore, the FRRs based on these two approaches are essentially \textit{different variants}
of the same formal construct, i.e. different FRRs, since they lead to expressions for the response functions involving different correlation functions in the absence of the force.
It can be interesting to discuss, what are general conditions making possible the elimination of explicit time dependence by augmenting the state space.   

\bigskip

\section{Conclusions and outlook.}
In the present work we provided a simple and straightforward derivation of a class of FRRs for non-stationary processes, within the line of reasoning mostly close to the
one used for derivation of FRRs in  equilibrium and NESS cases. The main difference of these FRRs from the customary ones is
the unusual nature of the variable conjugate to the force (dynamic conjugate): This variable depends on the values of internal variables at two different times. If the system of interest is prepared 
in the stationary state,
the $F_-$ variant of our FRR coincides with the customary FRR in equilibrium or in NESS.

The main property of the dynamics important for our derivation is the Markovianity, i.e., Eqs.~(\ref{eq:CInt}) and (\ref{eq:CK0}). The only other important
property is the differentiability of the transition PDF with respect to parameters constituting the forcing vector $\vec{F}$.
No other requirements are crucial. Like in \cite{Verley}, we do not assume the force-free dynamics to be autonomous, i.e., can consider explicitly time-dependent Markovian situations.  
Moreover, we do not assume the densities $p(... ;\{ F \})$ to be proper PDFs normalized to unity, if only  these densities satisfy Eqs.~(\ref{eq:CInt}) and (\ref{eq:CK0}), and can be used to obtain the means of $\mathcal{O}(\mathbf{x})$ defined correspondingly. This allows to apply the formalism to improper cases (absorbing states, sources, or birth-death processes), cf. \cite{Padmanabha}.
A set of simple examples elucidates some properties of static and dynamic conjugates, and the relations between them. The discussion of further generalizations leading to 
more interesting physics is planned for a forthcoming work.

\appendix

\section{Explicit calculations pertinent to Ornstein-Uhlenbeck example \label{Sec:OU}}

The OU example is instructive, because we essentially know the answer, and it is interesting to see how this answer is represented in different variants of a discussion we have conducted.
The transition probability in the Ornstein-Uhlenbeck process described by the Langevin equation
\begin{equation}
m \dot{v} = - \frac{1}{\mu} v + F + \sqrt{\frac{2 k_B T}{\mu}} \xi(t),
\label{eq:OUO}
\end{equation}
with $m$ being the particle's mass, $\mu$ its mobility, and $\xi(t)$ being the standard Gaussian white noise with $\langle \xi(t) \rangle = 0$, $\langle \xi(t) \xi(t)\rangle = \delta(t-t')$, is given by
\begin{eqnarray}
 && p(v,t|v',t';F) = \frac{1}{\sqrt{2 \pi \langle v^2 \rangle_{eq} [1-e^{-\frac{2(t-t')}{\tau}}]}} \label{eq:ConDis}\\
 && \times \exp \left\{-\frac{1}{2\langle v^2 \rangle_{eq}} \frac{[v-v'e^{-\frac{t-t'}{\tau}} - \mu F(1-e^{-\frac{t-t'}{\tau}})]^2}{1-e^{-\frac{2(t-t')}{\tau}}} \right\} \nonumber
\end{eqnarray}
with $\langle v^2 \rangle_{eq} = \frac{k_B T}{m}$ and $\tau = \mu m$. To obtain the result it is enough to note that introducing the variable 
$u = v - \mu F$ (the deviation of velocity from the terminal one) reduces Eq.~(\ref{eq:OUO}) to the genuine Ornstein-Uhlenbeck one \cite{OU}, for which the original formula for the 
transition PDF applies. Note that the transformation must be performed both for the velocity at running time $t$ and for the initial velocity $v'$ at time $t'$.  
From Eq.~(\ref{eq:ConDis}) we get:
\begin{eqnarray}
 \mathcal{Y}(v,t;v',t') &=& \mu \frac{1}{V^2} \frac{1-e^{-\frac{t-t'}{\tau}}}{1-e^{-\frac{2(t-t')}{\tau}}} \left(v - v' e^{-\frac{t-t'}{\tau}} \right) \nonumber \\
 &=& \frac{\mu m}{k_B T} \frac{v-v' e^{-\frac{t-t'}{\mu m}}}{1+e^{-\frac{t-t'}{\mu m}}}. \label{eq:TranPro}
\end{eqnarray}
This variable is explicitly time dependent. The expression above is valid for any $t' < t$, and can be used also for obtaining 
\[
 \mathcal{Y}(v',t';v_0,t_0) =  \frac{\mu m}{k_B T} \frac{v'-v_0 e^{-\frac{t'-t_0}{\mu m}}}{1+e^{-\frac{t'-t_0}{\mu m}}},
\]
necessary for the discussion of the $F_-$ setup. In the $F_-$ case, the limiting value for $t_0 \to - \infty$ is 
\begin{equation}
Y(v') = \lim_{t_0 \to - \infty} \mathcal{Y}(v',t';v_0,t_0) =  \frac{\mu m}{k_B T} v'.
\label{eq:YOU}
\end{equation}
Let us however consider the $F_+$ setup, take $v(t)$ as a VoI, and consider the cases when we start from an equilibrium state, and when we start with a well-prescribed velocity $v_0$. 
From Eq.~(10) of the main text it follows that
\begin{eqnarray*}
&& \langle v(t) | F_+ \rangle - \langle v(t) | 0 \rangle = \\
&& \frac{\mu m F}{k_B T(1+e^{-\frac{t-t'}{\mu m}})} \left[\langle v^2(t) \rangle - \langle v(t) v(t') \rangle e^{-\frac{t-t'}{\mu m}} \right],
\label{eq:VelNS}
\end{eqnarray*}
where the r.h.s. is calculated for given initial conditions but without perturbation. When starting from the equilibrium state, we have
\[
\langle v(t) | 0 \rangle =0, \;\; \langle v^2(t)\rangle = \frac{k_B T}{m}, \;\;  \langle v(t) v(t')  \rangle = \frac{k_B T}{m}  e^{-\frac{t-t'}{\mu m}},
\]
so that
\begin{eqnarray*}
&& \langle v(t) | F_+ \rangle - \langle v(t) | 0 \rangle =  \\
&&\qquad \mu F_+ \frac{1-e^{-\frac{2(t-t')}{\mu m}}}{1+e^{-\frac{t-t'}{\mu m}}}  = \mu F_+ [1-e^{-\frac{t-t'}{\mu m} }],
\end{eqnarray*}
giving the response function 
\begin{equation}
 \chi(t,t',0) = m^{-1} e^{-\frac{t-t'}{\mu m}}. \label{eq:ResF}
\end{equation}

More interesting is to see what happens if we start from a well-prescribed velocity $v_0 \neq 0$.
In this case 
\begin{equation}
 \langle v(t) | 0 \rangle = v_0 e^{-\frac{t}{\mu m}}, \label{eq:Mv0}
\end{equation}
and both $\langle v^2(t) \rangle$ and $\langle v(t) v(t') \rangle$ have to be calculated for a given initial condition. 
Since we know the transition probability, which in this case ($F=0$) is
\begin{eqnarray*}
 p(v,t|v',t') = && \sqrt{\frac{m}{2 \pi k_B T \left[1-e^{-\frac{2(t-t')}{\mu m}} \right] } } \\
&& \times \exp \left\{-\frac{m}{2 k_B T} \frac{\left[v-v'e^{-\frac{t-t'}{\mu m}}\right]^2}{1-e^{-\frac{2(t-t')}{\mu m}}} \right\}
\end{eqnarray*}
we get
\begin{eqnarray}
 \langle v^2(t) \rangle &=& \int v^2 p(v,t|v_0,0) dv \nonumber \\
 &=& \frac{k_B T}{m} (1-e^{-\frac{2t}{\mu m}}) + v_0^2 e^{-\frac{2t}{\mu m}} \label{eq:vsquared}
\end{eqnarray}
and
\begin{eqnarray}
  \langle v(t) v(t') \rangle &=& \int \int v v' p(v,t|v',t') p(v',t'|v_0,0) dv dv' \nonumber \\
  &=& e^{-\frac{t-t'}{\mu m}} \left[k_B T(1-e^{-\frac{2t'}{\mu m}}) + v_0^2 e^{-\frac{2t'}{\mu m}} \right] . \label{eq:CFv0}
\end{eqnarray}
To obtain the last expression we first integrate over $v$ and use the fact that $\langle v(t) | v'(t') \rangle = \int v p(v,t|v',t') dv = v' e^{-(t-t')/\mu m}$, and then resort to Eq.~(\ref{eq:vsquared}),
where $v$ and $t$ are changed for $v'$ and $t'$, respectively. Substituting the last two expressions into Eq.~(10) of the main text  we get:
\[
  \langle v(t) | F_+ \rangle - \langle v(t) | 0 \rangle =  \mu F [1-e^{-\frac{t-t'}{\mu m}} ],
\]
exactly as in the equilibrium case. In spite of the fact that we have started from a non-equilibrium condition, and the process $v(t)$ is now non-stationary, we finally get the same response function. 
As follows from our calculations above, this fact is due to exponential relaxation of all relevant properties, and does not have to hold for non-exponential cases. 

Similar calculations can be performed in the $F_-$ setup under given initial conditions. Turning to the stationary case (with $t_0 \to -\infty$), we get, using Eq.~(\ref{eq:YOU}), the standard, equilibrium FRR
\begin{eqnarray*}
 \langle v(t) | F_- \rangle - \langle v(t) | 0 \rangle &=& F \langle v(t) Y(t') \rangle_0 =F \frac{\mu m}{k_B T} \langle v(t) v(t') \rangle_0  \\
 &=&  \mu F e^{-\frac{t-t'}{\mu m}}
\end{eqnarray*}
with $\langle v(t) | 0 \rangle = 0$, leading to the same response function. 

The $\mathcal{Z}$- (Verley, Ch\'etrite, and Lacoste's, \cite{Verley}) conjugate is obtained from 
\[
 p_1(v',t',t_0; F) = \int dv_0 p(v',t'|v_0,t_0;F) p(v_0)
\]
with $p(v',t'|v_0,t_0;F)$ given by Eq.~(\ref{eq:ConDis}).

In the equilibrium case, starting with a Maxwell distribution with $\langle v \rangle =0$
\[
 p(v_0) = \sqrt{\frac{m}{2 \pi k_B T}} \exp\left(- \frac{m v_0^2}{2 k_B T} \right)
\]
we have
\[
 \mathcal{Z}(v',t';t_0) = \frac{\mu m}{k_B T} \left(1- e^{-\frac{t'-t_0}{\tau}}\right) v',
\]
tending to $Y$ for $t_0 \to -\infty$. 
Putting $t_0 = 0$ we thus get:
\begin{eqnarray*}
&& \langle v(t) | F_- \rangle - \langle v(t) | 0 \rangle = F \langle v(t) \mathcal{Z}(v',t';t_0) \rangle_0  \\
&& \qquad F \frac{\mu m}{k_B T} \left(1- e^{-\frac{t'}{\mu m}}\right) \langle v(t) v(t') \rangle_0
\end{eqnarray*}
where the last mean is calculated for equilibrium initial conditions, and without forcing. This one is equal to 
$\langle v(t) v(t')  \rangle = \frac{k_B T}{m}  e^{-\frac{t-t'}{\mu m}}$, so that
\[
 \langle v(t) | F_- \rangle - \langle v(t) | 0 \rangle = \mu F \left(e^{-\frac{t-t'}{\mu m}} - e^{-\frac{t}{\mu m}} \right),  
\]
giving the same response function as given by Eq.~(\ref{eq:ResF}).

For the nonequilibrium case with sharp $v_0 \neq 0$, putting $t_0 = 0$ we get:
\[
 \mathcal{Z}(v',t';t_0) = \frac{\mu m}{k_B T} \frac{e^{t'/\tau} v' - v_0}{1+e^{t'/\tau}}.
\]
Therefore,
\[
\langle v(t) | F_- \rangle - \langle v(t) | 0 \rangle = F \frac{\mu m}{k_B T} \frac{e^{t'/\tau} \langle v(t') v(t) \rangle - v_0 \langle v(t) \rangle}{1+e^{t'/\tau}}
\]
Using Eqs.~(\ref{eq:Mv0}) and (\ref{eq:CFv0}) we get:
\[
 \langle v(t) | F_- \rangle - \langle v(t) | 0 \rangle = \mu F \left(e^{-\frac{t-t'}{\mu m}} - e^{-\frac{t}{\mu m}} \right),  
\]
like in the previous case. Note, however, that the free dynamics with $\langle v(t) | 0 \rangle$ given by Eq.~(\ref{eq:Mv0}) is now different from the one with $\langle v(t) | 0 \rangle=0$ in the previous case. 

\section{A non-proper case \label{App:Nonprop}}

The diffusion equation in presence of the constant force reads:
\begin{equation}
\frac{\partial w}{\partial t} = - \mu F \frac{\partial w}{\partial x} +D \frac{\partial^2 w}{\partial x^2} 
\label{eq:Drift1} 
\end{equation}
with $D$ being the diffusion coefficient, $\mu$ the mobility of the particle, and $F$ being the external force.
Denoting $f= - \mu F$ we can write the equation as 
\[
 \frac{\partial w}{\partial t} = f \frac{\partial w}{\partial x} +D \frac{\partial^2 w}{\partial x^2}. 
\]
The solution follows from a general discussion (\cite{Polyanin}, Sec. 1.1.4.). The change of variables 
\begin{equation}
 w(x,t) = \exp(a t + b x) u(x,t)
 \label{eq:ChanV}
\end{equation}
with 
\begin{eqnarray*}
a &=& - \frac{f^2}{4D} \equiv - \frac{\mu^2 F^2}{4 D} \\
b &=& - \frac{f}{2D} \equiv \frac{\mu F}{2 D} 
\end{eqnarray*}
reduces Eq.~(\ref{eq:Drift1}) to the heat equation (see Sec. 1.1.4-2 of \cite{Polyanin}) for $u$:
\begin{equation}
 \frac{\partial u}{\partial t} = D \frac{\partial^2 u}{\partial x^2}.
 \label{eq:u}
\end{equation}
Note that the initial condition $w(x,0) = \delta(x-x_0)$ for $w$ is transformed into the initial condition 
\[
 \delta(x-x_0) = \exp \left( \frac{\mu F}{2 D} x \right) u(x,0),
\]
for $u(x,t)$, from which it follows that
\begin{eqnarray}
 u(x,0) &=& \exp \left( - \frac{\mu F}{2 D} x \right)  \delta(x-x_0) \nonumber \\ 
 &\equiv& \exp \left( - \frac{\mu F}{2 D} x_0 \right)  \delta(x-x_0) .
\label{eq:init}
\end{eqnarray}

The reason for considering an interval with two absorbing boundaries as the simplest example is that the Dirichlet boundary condition with $w = 0$ is mapped to 
the same Dirichlet boundary condition for $u$, while the Neumann boundary condition which would apply at a reflecting boundary at $x=l$ would follow 
by differentiation of both parts of Eq.~(\ref{eq:ChanV}) w.r.t. $x$
\[
0 = \frac{\partial w}{\partial x} = b \exp(a t + b x) u(x,t) + \exp(a t + b x) \frac{\partial u}{\partial x},
\]
and is transformed to a radiative (Robin) boundary condition for $u$:
\[
 b u(l,t) + \left.\frac{\partial u(x,t)}{\partial x} \right|_{x=l} = 0.
\]
The solution for $u$ in this case can be obtained by taking a limit $k_1 \to \infty$ in Eq.~(1.1.1-11) of Ref.~ \cite{Polyanin}, but does not have the simplicity of the 
case considered here.

The solution of Eq.~(\ref{eq:u}) for the case $u(0) =0$, $u(l) =0$ with the initial condition $u(x,0) = \delta(x-x_0)$ is given by the Green's function
of the diffusion equation on an interval with two absorbing boundaries
\[
 G(x,x_0,t) = \frac{2}{l}\sum_{n=1}^\infty \sin\left(\frac{n \pi x}{l}\right) \sin\left(\frac{n \pi x_0}{l}\right) e^{-\frac{D n^2 \pi^2 t}{l^2}}.
\]
Note that in our (really simplest) case $G(x,x_0,t)$ is independent on $f$. Thus, in initial variables (and for the initial condition given by Eq.~(\ref{eq:init})) our solution reads:
\begin{eqnarray*}
 w(x,t|x_0,0) = && \exp \left(- \frac{\mu^2 F^2}{4 D} t + \frac{\mu F}{2 D} x \right) \\
 && \qquad \times \exp \left( - \frac{\mu F}{2 D} x_0 \right) G(x,x_0,t),
\end{eqnarray*}
which is Eq.~(\ref{eq:Nonp1}) of the main text. \bigskip

\section{Underdamped case \label{Sec:AppUnder}}.

We start from Eq.~(\ref{eq:FRRKK}) of the main text, whose r.h.s. contains the expression 
\[ \frac{1}{\alpha} \langle  \Delta x \Delta v \rangle + \langle \Delta x^2 \rangle 
\]
where the means are taken in the absence of the force. 
To avoid direct calculations of the means we note that in order to calculate $\langle \Delta x^2 \rangle$ we only need the marginal distribution of $\Delta x$, and the result can be obtained immediately:
\[
 \langle \Delta x^2 \rangle = \sigma_x^2 + \frac{q^2(t)}{\alpha^2} v_0^2.
\]
To calculate the first mean we note that  $\Delta x(t) = \int_0^t v(t') dt'$, and therefore
 \[
 \langle \Delta x(t) v(t) \rangle = \int_0^t \langle v(t') v(t) \rangle dt' .
\]
while 
\[
  \langle \Delta x(t) v(0) \rangle = v_0 \int_0^t \langle v(t') \rangle dt' = \frac{v_0^2}{\alpha}(1-e^{-\alpha t}) =  \frac{v_0^2 q(t)}{\alpha}.
\]
The (non-centered) velocity-velocity correlation function $\langle v(t') v(t) \rangle$ is given by Eq.~(\ref{eq:CFv0}), and in the recent notation reads
\[
 \langle v(t') v(t) \rangle = e^{-\alpha(t-t')} \left[v_{eq}^2 (1-e^{-2 \alpha t'}) + v_0^2 e^{-2 \alpha t'}\right].
\]
Thus,

\begin{equation}
 \langle \Delta x(t) v(t) \rangle = v_0^2 e^{-\alpha t} \frac{1-e^{-\alpha t} }{\alpha} +\frac{(1-e^{-\alpha t} )^2}{\alpha} v_{eq}^2. 
 \label{eq:Deltaxv}
\end{equation}
On the total, we get:
\begin{eqnarray*}
  \frac{1}{\alpha} \langle  \Delta x \Delta v \rangle + \langle \Delta x^2 \rangle &=& \frac{2 v_{eq}^2}{\alpha^2}\left[e^{-\alpha t} + t\alpha - 1 \right] \\
  &\equiv& \frac{2 k_B T}{\gamma} \left[t-\frac{1-e^{-\alpha t}}{\alpha} \right] .
\end{eqnarray*}
The r.h.s. of Eq.~(\ref{eq:FRRKK}) reads:
\begin{eqnarray*}
 \frac{F}{2 k_B T} \left( \frac{1}{\alpha} \langle  \Delta x \Delta v \rangle + \langle \Delta x^2 \rangle \right) = \frac{F}{\gamma} \left[t-\frac{1-e^{-\alpha t}}{\alpha} \right], 
\end{eqnarray*}
which can be confirmed by a direct calculation.


\begin{thebibliography}{99}

\bibitem{Sato} K. Sato, Y. Ito, T. Yomo, and K. Kaneko, On the relation between fluctuation and response in biological systems, PNAS \textbf{100}, 14086 –14090 (2003)
\bibitem{Yan} C.-C. Sanders Yan; C.-P. Hsu, The fluctuation-dissipation theorem for stochastic kinetics -- Implications on genetic regulations,  J. Chem. Phys. \textbf{139}, 224109 (2013)
\bibitem{Droste} F. Droste and B. Lindner, Exact analytical results for integrate-and-fire neurons driven by excitatory shot noise, J Comput Neurosci. \textbf{43}, 81-91 (2017)  
\bibitem{Lindner} B. Lindner, Fluctuation-Dissipation Relations for Spiking Neurons, Phys. Rev. Lett. \textbf{129}, 198101 (2022)
\bibitem{Kubo} R. Kubo, The fluctuation–dissipation theorem, Rep. Prog. Phys. \textbf{29}, 255 - 284 (1966)
\bibitem{Villamaina} D. Villamaina, A. Baldassarri, A. Puglisi, and A. Vulpiani, The fluctuation-dissipation relation:
how does one compare correlation functions and responses? J. Stat. Mech. P07024 (2009) 
\bibitem{Sevick} E.M. Sevick, R. Prabhakar, S. R. Williams, and D. J. Searles, Fluctuation Theorems, Annu. Rev. Phys. Chem. \textbf{59}, 603–633 (2008)
\bibitem{Darrigol} O. Darrigol, A history of the relation between fluctuation and dissipation, Eur. Phys. J. H  \textbf{48}, 10 (2023)
\bibitem{Vulpiani} U. M. B. Marconia, A. Puglisi, L. Rondoni and A. Vulpiani, Fluctuation–dissipation: Response theory in statistical physics, Physics Reports \textbf{461} 111–195 (2008) 
\bibitem{Ewa} E. Gudowska-Nowak, F. A. Oliveira, and H. S. Wio, Editorial: The Fluctuation-Dissipation Theorem Today, Front. Phys. \textbf{10}, 859799 (2022)
\bibitem{Baldovin} M. Baldovin, L. Caprini, A. Puglisi, A. Sarracino, and A. Vulpiani, The Many Faces of Fluctuation-Dissipation Relations Out of Equilibrium. In: Brenig, L., Brilliantov, N., Tlidi, M. (eds) 
Nonequilibrium Thermodynamics and Fluctuation Kinetics. Fundamental Theories of Physics, vol 208. Springer, Cham, 2022
\bibitem{Ewa_new} M.S. Gomes-Filho, L.C. Lapas, E. Gudowska-Nowak, F.A. Oliveira, The fluctuation–dissipation relations: Growth, diffusion, and beyond, Physics Reports \textbf{1141}, 1-43 (2025)
\bibitem{Sarracino} A. Sarracino and A. Vulpiani, On the fluctuation-dissipation relation in non-equilibrium and non-Hamiltonian systems,  Chaos \textbf{29}, 083132 (2019)
\bibitem{Seifert_2}  U. Seifert, Stochastic thermodynamics, fluctuation theorems and molecular machines,” Rep. Prog. Phys. \textbf{75}, 126001 (2012).
\bibitem{Baiesi} M. Baiesi and C. Maes, An update on the nonequilibrium linear response, New J. Phys. \textbf{15}, 013004 (2013)
\bibitem{Goerlich} R. Goerlich, A. Tartar, Y. Roichman, and I.M. Sokolov, Fluctuation-response relation for a nonequilibrium system with resolved Markovian embedding, Phys. Rev. E, \textbf{114}, 014138 (2026)
\bibitem{OMa} L. Onsager and S. Machlup, Fluctuations and Irreversible Processes, Phys. Rev. \textbf{91}, 1505 - 1512 (1953)
\bibitem{Willareth} L. Willareth, I. M. Sokolov, Y. Roichman, and B. Lindner, Generalized fluctuation-dissipation theorem as a test of the Markovianity of a system, EPL \textbf{118}, 20001 (2017)
\bibitem{Engbring} K. Engbring, D. Boriskovsky, Y. Roichman, B. Lindner, A Nonlinear Fluctuation-Dissipation Test for Markovian Systems, Phys. Rev. X \textbf{13}, 021034 (2023)
\bibitem{Podhaisky} G. Podhaisky, I.M. Sokolov, Y. Roichman, and B. Lindner, Reliability of a nonlinear fluctuation-dissipation relation as a test of Markovianity, Phys. Rev. E, \textbf{113}, 044130 (2026)
\bibitem{Caprini} L. Caprini, A. Puglisi, and A. Sarracino, Fluctuation–Dissipation Relations in Active Matter Systems, Symmetry \textbf{13}, 81 (2021)
\bibitem{Seifert} U. Seifert and T. Speck, Fluctuation-dissipation theorem in nonequilibrium steady states, EPL \textbf{89}, 10007 (2010)
\bibitem{van_Kampen} N. G. van Kampen, Stochastic Processes in Physics and Chemistry, 3rd ed. (North Holland, Amsterdam, 2007) pp. 73–79.
\bibitem{Feller_ex} W. Feller, Non-Markovian processes with semigroup property, Ann. Math. Statist. \textbf{30}, 1252 (1959)
\bibitem{Canturk} B. Canturk and H.-P. Breuer, On positively divisible non-Markovian processes, J. Phys. A: Math. Theor. \textbf{57} 265006  (2024)
\bibitem{Wiki} \url{https://en.wikipedia.org/w/index.php?title=Fluctuation%E2%80%93dissipation_theorem&oldid=1279400208}, accessed on September 4, 2026
\bibitem{Lenk} R. Lenk, A simple proof of the classical fluctuation dissipation theorem, Physics Lett., \textbf{25A}, 198 - 199 (1967)
\bibitem{GreeneCallen1} R.F. Greene and H.B. Callen, On the Formalism of Thermodynamic Fluctuation Theory, Phys. Rev \textbf{83}, 1231 - 1235 (1951)
\bibitem{Takahasi} H. Takahasi, Generalized Theory of Thermal Fluctuations, J. Phys. Soc. Japan, \textbf{7}, 439 - 446 (1952)
\bibitem{Sokolov} I.M. Sokolov, Linear Response and Fluctuation-Dissipation Relations for Brownian Motion under Resetting, Phys. Rev. Lett. \textbf{130}, 067101 (2023)
\bibitem{Prost} J. Prost, J.-F. Joanny, and J.M.R. Parrondo, Generalized Fluctuation-Dissipation Theorem for Steady-State Systems, Phys. Rev. Lett. \textbf{103}, 090601 (2009) 
\bibitem{Sasa} T. Hatano and S. Sasa, Steady-State Thermodynamics of Langevin Systems, Phys. Rev. Lett. \textbf{86}, 3463 (2001)
\bibitem{Altaner} B. Altaner, M. Polettini, and M. Esposito, Fluctuation-Dissipation Relations Far from Equilibrium, Phys. Rev. Lett. \textbf{117}, 18601 (2016) 
\bibitem{Goerlich2} R. Goerlich, B. Sorkin, D. Boriskovsky, L. B. Pires, B. Lindner, C. Genet, and Y. Roichman, Consistent thermodynamics reconstructed from transitions between nonequilibrium steady-states,
arXiv:2601.03245.
\bibitem{EvaMaj} M.R. Evans and S.N. Majumdar, Diffusion with Stochastic Resetting, Phys. Rev. Lett. \textbf{106}, 160601 (2011).
\bibitem{Verley} G. Verley, R. Ch\'etrite, and D. Lacoste, Modified fluctuation-dissipation theorem for general non-stationary states and application to the
Glauber–Ising chain, J. Stat. Mech. P10025 (2011)
\bibitem{Padmanabha} P. Padmanabha, S. Azaele, and A. Maritan, Generalisation of fluctuation-dissipation theorem to systems with
absorbing states, New J. Phys. \textbf{25} 113001 (2023)
\bibitem{Polyanin} A.D. Polyanin, Handbook of Linear partial Differential Equations for Engineers and Scientisis, Chapman \& Hall/CRC Press, Boca Raton, 2002
\bibitem{OU} G.E. Uhlenbeck and L.S. Ornstein, On the Theory of the Brownian Motion, Phys. Rev. 36, 823 - 841 (1930)
\bibitem{Chandra} S. Chandrasekhar, Stochastic problems in physics and astronomy, Rev. Mod. Phys. \textbf{15}, 1 (1943)


\end{thebibliography}
\end{document}